\documentclass{pasj00}
\begin{document}
\SetRunningHead{Deguchi et al.}{SiO Masers Search off the Galactic Plane}
\Received{2006/12/18}
\Accepted{2007/03/27 ; Ver 2.2 March 28, 2007}

\title{An SiO Maser Search off the Galactic Plane}

\author{Shuji \textsc{Deguchi},$^{1}$  
        Takahiro \textsc{Fujii},$^{2,3}$ 
         Yoshifusa \textsc{Ita},$^{4,5}$, Hiroshi \textsc{Imai},$^{2,3}$,}
\author{Hideyuki \textsc{Izumiura},$^{6}$ Osamu \textsc{Kameya},$^{7}$ 
        Noriyuki \textsc{Matsunaga},$^{4}$ Atsushi \textsc{Miyazaki},$^{1,7,8}$} 
\author{Arihiro \textsc{Mizutani},$^{9,10}$ Yoshikazu \textsc{Nakada},$^{4}$ 
        Jun-ichi \textsc{Nakashima},$^{1,11}$}
\and 
\author{Anders \textsc{Winnberg}$^{1,12}$}

\affil{$^{1}$ Nobeyama Radio Observatory, National Astronomical Observatory,\\
              Minamimaki, Minamisaku, Nagano 384-1305}
\affil{$^{2}$ VERA Project Office, National Astronomical Observatory, \\
              2-21-1 Osawa, Mitaka, Tokyo 181-8588}
\affil{$^{3}$ Faculty of Science, Kagoshima University, 
              1-21-35 Korimoto, Kagoshima 890-0065}
\affil{$^{4}$ Institute of Astronomy, School of Science, The University of Tokyo,\\
              2-21-1 Osawa, Mitaka, Tokyo 181-0015}
\affil{$^{5}$ Institute of Space and Astronautical Science, Japan Aerospace Exploration Agency, \\
               Yoshinodai 3-1-1, Sagamihara, Kanagawa 229-8510}
\affil{$^{6}$ Okayama Astrophysical Observatory, National Astronomical 
              Observatory, \\ Kamogata, Asakuchi, Okayama 719-0232}
\affil{$^{7}$ Mizusawa VERA Observatory, National 
               Astronomical Observatory, \\ Mizusawa, Oshu, Iwate 023-0861}
\affil{$^{8}$ Shanghai Astronomical Observatory, Chinese Academy of Sciences \\
               80 Nandan Road, Shanghai, 200030, P.R. China}
\affil{$^{9}$ Department of Astronomical Science, The Graduate University 
               for Advanced Studies,\\ 2-21-1 Osawa, Mitaka, Tokyo 181-8588}   
\affil{$^{10}$ Koriyama City Fureai Science Center, 2-11-1 Ekimae, Koriyama, Fukushima 963-8002} 
\affil{$^{11}$ Academia Sinica, Institute of Astronomy and Astrophysics, \\
               P.O. Box 23-141, Taipei 106, Taiwan}
\affil{$^{12}$ Onsala Space Observatory, 439 92 Onsala, Sweden} 




%


\maketitle

\begin{abstract}

We have searched for the SiO $J=1$--0 $v=$ 1 and 2 maser lines at $\sim 43$ GHz
in 277 2MASS/MSX/IRAS sources off the Galactic plane ($|b|\gtrsim 3^{\circ}$),
which resulted in 119 (112 new) detections. Among the new detections,
are two very faint objects with MSX $12 \mu$m flux densities below 2 Jy. These 
 are likely to be O-rich AGB-stars associated with dwarf-galaxy tidal tails.
The sample also includes medium bright MSX objects at moderately high Galactic latitudes
($3^{\circ}<|b|<5^{\circ}$) and in the IRAS gap at higher latitudes.
A signature of a warp of the inner Galactic disk is found for a disk subsample. 
This warp appears relatively strongly in the area of $0<l<45^{\circ}$\ and $3<|b|<5^{\circ}$. 
We also found a group of stars that does not follow to the Galactic rotation. This feature appears   
in the Galactic disk at $l\sim 27^{\circ}$, and extends more than $15^{\circ}$ in Galactic latitude, 
like a stream of tidal debris from a dwarf galaxy.  

\end{abstract}

\section{Introduction}
Stellar maser sources are good tracers of Galactic structure. 
They are mostly AGB (Asymptotic Giant Branch) stars
with typical ages of a few Gyr, representing a dynamically well-relaxed component of the Galaxy.    
Large surveys of maser sources have been made using the OH 1612 MHz line in the past
towards objects in the Galactic plane within $|l|<45^{\circ}$ and $|b|<3^{\circ}$ \ \citep{sev97,sev01},
or using the 43 GHz ($J=1$--0 $v=1$ and 2) SiO maser lines toward the Galactic bulge \citep{izu95},  
inner and outer disk \citep{deg04b,jia96}, and the Galactic center \citep{deg04a,fuj06}, or
with the 86 GHz  ($J=2$--1 $v=1$) SiO maser line in a thinner strip $|b|\lesssim 1^{\circ}$ \citep{mes02}.
However, because of a relative scarcity of sources off the Galactic plane ($|b|>3^{\circ}$), all-sky surveys 
for the maser sources were not made much in the past (except for \cite{tel91} or \cite{ita01}).   
Because of recent studies of halo sub-structure \citep{iba01a} and it's dynamical 
influence on the disk structure of the Galaxy \citep{bin01}, 
it is worth looking among SiO maser survey data for signatures 
of dwarf galaxies in the Galactic halo. 
In fact, the MSX survey, which was a large deep survey 
of infrared objects after IRAS, contained data from the Galactic disk
($|b|<6^{\circ}$) and additional higher-latitude objects 
in the limited area that IRAS could not access (the "IRAS gap"). 
Many of these higher-latitude objects were not searched for masers in the past simply because
they were absent from the IRAS Point Source Catalog, even though they are bright
in the infrared.         

In this paper, we report on the results of SiO maser searches for
objects outside the Galactic plane. The sample contains very bright to unusually 
faint IR sources, because they originate from several observation programs with different rationales.
Some of them are very distant objects for which SiO masers were never thought to be detectable, 
and therefore interesting in themselves.

\section{Observations}

Simultaneous observations in the SiO $J=1$--0, $v=1$ and 2 transitions at
42.122 and 42.821 GHz respectively were made with the 45m radio telescope
at Nobeyama mainly during the winter--spring seasons between 2003 --2006: 
in addition a few unpublished results from 1999 and 2000 were included in this work.    
We used a cooled SIS mixer receiver (S40) for the 43 GHz
observations and accousto-optical spectrometer arrays, AOS-H and AOS-W,
having bandwidths of 40 and 250 MHz respectively. The effective velocity resolution of
the AOS-H spectrometer was 0.3 km s$^{-1}$. The arrays covered the velocity range of
$\pm 380 $ km s$^{-1}$ (except for the 2006 observations which had a $\pm 150$ km s$^{-1}$ coverage), 
for both the SiO $J=1$--0 $v=1$ and 2 transitions
simultaneously. The overall system temperature was between 200 and 300 K,
depending on weather conditions. The half-power telescope beam width
(HPBW) was about 40$''$. The antenna temperature 
is corrected for atmospheric and telescope ohmic loss, but not for the beam or aperture efficiency
($T_{\rm a}^{*}$).
The factor for conversion between antenna temperature and flux density is 
about 2.9 Jy K$^{-1}$. Further
details of SiO maser observations using the NRO 45-m telescope have been
described elsewhere (\cite{deg00a}).

The objects in the present work originate from several different observing
programs, which were partially backup programs during times when the weather
was unsuitable for higher frequency observations.
Therefore, results from several different categories of objects are 
presented here. A majority of our objects are
MSX sources off the Galactic plane in the area, $0^{\circ}<l<45^{\circ}$ 
and $3^{\circ}<|b|<6^{\circ}$; these objects
are mostly excluded from our previous systematic SiO surveys of the Galactic plane,
which were limited to the strip with $|b|<3^{\circ}$ \citep{deg04b}. A second group of objects
are MIR sources from the $|b|>6^{\circ}$ MSX catalog \citep{ega03}
in the IRAS gap; these are often bright MIR objects with no IRAS record and 
as a result they have not been thoroughly investigated before. The third group of objects are
very faint stars which were chosen for the search for AGB stars 
from halo dwarf galaxies. These objects are mostly located at high 
Galactic latitudes; they exhibit star-like red images on 2MASS $JHK$-band images,
but the AGB nature of these objects has not been well established due to their faintness in the infrared bands. 
These objects were selected by customary color-index criteria for SiO maser searches \citep{deg04b}:
$H-K>0.5$, and $-0.5 \lesssim C_{CE} \lesssim 0.5$. Here $C_{CE}=log(F_{E}/F_{C})$,
and $F_{C}$ and $F_{E}$ are the MSX flux densities in the $C$(12 $\mu$m) and $E$(21 $\mu$m) bands.  
The $K$ magnitude and the 12 $\mu$m flux density are measures of distance to the source.
On average, the mass losing AGB stars at about 8 kpc (as for stars  in the Galactic bulge) typically have 
a color-corrected $K$ magnitude of $\sim 6$ and a 12 $\mu$m flux density of $\sim 4$ Jy
(for objects with an effective dust-shell temperature of 300 K). Therefore, when searching distant objects,
like those of Sgr Dwarf galaxy and it's tail ($D\sim 16$--24 kpc), we have to look for faint objects 
with $K\gtrsim 8$ and $F_C\sim 1$ Jy or less. For the case of CMa dwarf ($D\sim 7$ kpc),
the flux limitation is slightly relaxed. In order to search for SiO masers in such a distant object,
we have to observe all the brighter objects in the same area to get a good idea of the velocity field 
of foreground Galactic sources. Thus, we included the brighter objects in front of 
the dwarf galaxies in the survey, especially for the bulge sources toward Sgr dwarf galaxy 
(center at $l=5^{\circ}$\ and $b=-14^{\circ}$; \cite{maj03}).
For the case of the CMa dwarf (center at $l=240^{\circ}$\ and $b=-8^{\circ}$ ; \cite{mar04b}),
it is problematic too because the distance is around 7 kpc and some of objects spread toward the Galactic disk. 
For the survey, we chose faint red objects within 30 degree of the center of the CMa dwarf.
    
We searched for the SiO $J=1$--0 $v=1$ and 2 maser lines in 277 objects and detected 119 (112 new) sources.  
The detected sources are summarized in table 1 and the undetected sources in table 2. 
The distribution of the observed sources on the sky is shown in figure 1. 
Table 3 lists the IR properties of the observed sources: 2MASS name, 2MASS $K$ magnitude, $J-H$ 
and $H-K$ color indices,
MSX 6C name, position differences between 2MASS and MSX sources, MSX C-band (12$\mu$m) flux densities,
$C_{AC}=log(F_C/F_A)$ and $C_{CE}=log(F_E/F_C)$, and IRAS name and its' separation from the 2MASS position.
The 2MASS $K$ magnitude and MSX C-band (12 $\mu$m) flux density are plotted against 
color index of the observed objects in the left and right panels of figure 2, respectively. 
Though the observed objects involved a wide range of flux densities in terms of 2MASS $K$ or MSX C band
compared with those of other well defined survey programs,
the color indices of the objects were limited to a reasonable range for 
SiO detections \citep{deg00a,deg04b,fuj06}. 
Our sample involves 25 AGB candidates for dwarf galaxy associations, which resulted in 3 SiO detections. 
We discuss individual objects separately in Appendix A and provide the 
112 spectra of our new detections there. 


\section{Discussion}

The observed radial velocities are plotted against Galactic longitude in figure
3. Velocities expected from Galactic rotation are shown by solid and broken
curves for various galactocentric distances. The data have been divided into two
sub-samples: one `disk sample' with $|b|<10^{\circ}$ and one `high-latitude
sample' with $|b|>10^{\circ}$. There is a hint in the upper panel of figure 3
that the disk sample (open circles) follows the Galactic rotation curve better
than the high-latitude sample (filled triangles).  Most of the high-latitude
objects are supposedly Galactic bulge stars, so deviations of up to
about 100~km~s$^{-1}$ from the Galactic rotation are to be expected \citep{izu95}.  

\subsection{AGB candidates for dwarf galaxy associations}
The Sgr dwarf elliptical galaxy (SagDEG) was discovered by \citet{iba94} 
at $V_{hel}=140$ km s$^{-1}$ toward the Galactic bulge. 
Its distance was estimated to be approximately 24 kpc from the Sun.
Later investigations have confirmed the associations of M54 and other globular clusters with this dwarf galaxy, 
and moreover its associated trail of stars along its orbit being almost perpendicular to the Galactic plane
\citep{iba01b}. The trail includes a number of C-rich AGB stars \citep{iba01a} and M giants \citep{maj03}.
Therefore, the idea of looking for O-rich mass-losing AGB stars is not far-fetched, 
though the low-metal low-mass stars
tend to become carbon-rich stars at the AGB phase of stellar evolution \citep{ibe81,iba01b}. 
In fact, an SiO maser source have been detected 
in the Large Magellanic cloud with the SEST 15m telescope at 86.2 GHz (\cite{van96};
see also \cite{woo92} for OH/IR sources).  
 
An enhanced stellar density was found toward the anticenter region in the Sloan Digital Sky Survey \citep{yan03}, 
suggesting that this is also a remnant of adwarf galaxy. The center of this enhanced density was estimated to be 
$l=240$--244$^{\circ}$, $b=-8$ -- $-10^{\circ}$, and this was named
as CMa dwarf galaxy \citep{bel04,mar04a}. Because the distance and radial velocity of the CMa dwarf galaxy are estimated 
to be $7.2\pm 0.3$ kpc and $109\pm 4$ km s$^{-1}$ respectively \citep{mar04b}, 
our search for SiO masers in this object may be slightly easier than is the case in the  
Sgr dwarf galaxy. 

We detected three of the 25 surveyed AGB candidates for the dwarf-galaxy associations.   
Properties of these candidates are summarized in table 4, 
which gives the candidate group affiliation (Sgr or CMa candidates),
radial velocity in the Galactic Standard of Rest $V_{GSR}$, $\Lambda_{\odot}$ and $\beta_{\odot}$ 
(the SagDEG coordinates), 
$K$ and $H-K$, interstellar extinction $A_K$, estimated distance, IRAS 12 $\mu$m flux density,
and IRAS name. Here the distance was estimated from the difference between 
$K_c$ (interstellar- and circumstellar-extinction corrected $K$ magnitude) 
and a standard $m_K=6.43$ at the Galactic center
(K magnitude for Miras without circumstellar extinction with light variation period of 450 days 
at 8 kpc; see equation (5) of \cite{gla95}). 
Here, the corrected $K$ magnitude, an indicator of distance modulus which is corrected 
for the interstellar and circumstellar reddening (\cite{fuj06}), is defined by  
\begin{equation}
K_{c}= K- A_K/E(H-K) \times [(H-K)-(H-K)_0]
\end{equation}
and we use $A_K/E(H-K)=1.44$ \citep{nis06}, and $(K-H)_0=0.5$ (\cite{fuj06}).
Among the surveyed red objects, it turns out that $J$14572697$+$0516034 is a carbon star \citep{mau05}. 
Since the associations with the dwarf galaxies are not very certain, 
we discuss the three SiO detected objects individually. 

\begin{itemize}
\item $J$05475868$-$3305109 (IRAS 05461$-$3306): This is a faint IRAS source  with $F_{12}$= 2 Jy. 
The 2MASS $K$ magnitude is 6.15, suggesting its distance is about 7 kpc. 
The high radial velocity ($V_{\rm LSR}=110$ km s$^{-1}$)
and location in the sky ($b=-26.9^{\circ}$) suggest that this is associated with the CMa dwarf.

\item $J$08144298$-$3233047 (=IRAS 08127$-$3223): This is a medium bright MSX source G250.9283+01.2198 
with $F_{C}$= 9.4 Jy. Its 2MASS counterpart ($K=8.44$) has a red $H-K=2.17$.
The interstellar extinction toward this object is quite small ($A_K \sim 0.2$).
This object must be an O-rich AGB star undergoing copiously mass-loss. Because of its low galactic latitude
 ($b=1.2 ^{\circ}$) and small radial velocity ($V_{\rm LSR}=69$ km s$^{-1}$), this star is probably a disk star.
 
\item $J$19235554$-$1302029 (=IRAS 19211$-$1307): 
This object is a faint ($F_{12}$= 1.1 Jy) IRAS source at the edge of the Galactic bulge 
[($l$,$b$)=($24.8^{\circ}$, $-13.0^{\circ}$)].
The 2MASS $K$ magnitude ($K=9.7$) suggests that this star may be located at a distance 
more than 18 kpc from the Sun, and so could be associated to SagDEG or it's tail, 
though the radial velocity
($V_{\rm LSR}=-23.0$ km s$^{-1}$) does not fit the velocity of SagDEG.
\end{itemize}

Figure 4 shows the velocity--longitude diagram in the SagDEG coordinates for SiO sources;
the velocity given is based on the Galactic standard of rest (GSR) 
by adopting the maginude of the Galactic rotation
 being 220 km s$^{-1}$, and the longitude ($\Lambda _{\odot}$) is measured 
 along the orbital plane of  the Sgr dwarf galaxy, with 
the north pole at $l=273.8^{\circ}$ and $b=-13.5^{\circ}$ \citep{maj03}. By measuring the radial velocities of
M giants in the main trail of SagDEG, \citet{maj04} showed that  the SagDEG objects fall in a narrow strip crossing
at $\Lambda_{\odot} \sim 65^{\circ}$\ with $V_{GSR}=0$ km s$^{-1}$, 
and that the dispersion of radial velocities of the SagDEG trail is about 10 km s$^{-1}$. 
Our candidate star, $J$19235554$-$1302029 (=IRAS 19211$-$1307), which is located 
about 17$^{\circ}$\ below the Sgr orbital plane,
deviates from the expected velocity of the tidal stream of SagDEG ($V_{\rm SGR}\sim 160$ km s$^{-1}$) 
at this direction by more than 100 km s$^{-1}$.
However, the numerical simulation of the tidal debris (figure 10 of \cite{law05}) 
suggests that the trail of SagDEG can produce the $V\sim 50$ km s$^{-1}$ 
sub-streaming component in this direction;
the sub stream is indicated in figure 4 as a thin band between $\Lambda\sim 10^{\circ}$ and 
$150^{\circ}$  at $V_{\rm SGR}$ range between 0 and 180 km s$^{-1}$. Though the numerical simulations
\citep{iba01b,hel01} have considerable uncertainty in radial velocities of the sub-stream stars, which were
torn from SagDEG in ancient times, the orbits must cross near the present SagDEG position
in any simulations.  Therefore, the velocity of $V_{\rm GSR}=67$ km s$^{-1}$
at $\Lambda = 4.2^{\circ}$ does not exclude the possibility 
of this object being associated to the Sgr dwarf stream.   

Our observation shows that the radial velocity of $J$05475868$-$3305109, $V_{\rm LSR}=109$ km s$^{-1}$, 
coincides with the above-noted velocity of CMa dwarf \citep{mar04b,mar05}. The estimated distance
of about 7 kpc for this source (table 4) coincides with the estimated distance of CMa dwarf. 
These facts suggest an association of this object
to the CMa dwarf galaxy. However, $J$08144298$-$3233047 is unlikely to be 
associated to CMa dwarf because of its low radial velocity (68.8 km s$^{-1}$) 
and low Galactic latitude ($b=1.2^{\circ}$).

Even though the association of these objects with dwarf galaxies or their tails is somewhat uncertain, 
they are still O-rich mass-losing AGB stars located far away from the Sun, and as such are interesting objects.
The progenitors of O-rich AGB stars are considered to be more massive than those of C-rich AGB stars,
their presence in dwarf galaxies would constrain chemical-evolution and pollution models 
of dwarf galaxies (e.g., \cite{mcw05}).  

\subsection{A warp signature for the off-plane objects.}
It is interesting to check whether the velocity distribution 
of the off-plane objects ($3^{\circ}<|b|<5^{\circ}$) exhibits any differences 
from the lower-latitude ($|b|<3^{\circ}$) velocity distribution.
Upper panel of figure 5 shows a comparison of the velocity distribution 
between the "on-plane" (lower latitude) and off-plane samples: one from the SiO maser data
with $|b|<3^{\circ}$ and $0<l<45^{\circ}$ \citep{deg04b}; the other with $3<|b|<5^{\circ}$
and $0<l<45^{\circ}$ (from the present paper).
For the purpose of making the distance cut off as similar as possible for both samples, we
have chosen the 127 stars with $F_C> 6.5$ Jy within an area, $0<l<45^{\circ}$ and $|b|<3^{\circ}$
for the on-plane sample. At first glance, 
this diagram gives the impression that the off-plane sample contains a smaller number of
 objects with $V_{lsr}>100$ km s$^{-1}$ as  
compared with those in the on-plane sample. The average of $V_{lsr}$ 
for the off-plane sample (63 points) is 19.1 ($\pm 60.1$) km s$^{-1}$, while the average of  
$V_{lsr}$  for the on-plane sample  (127 points) is 40.6 ($\pm 59.8$) km s$^{-1}$.
This difference of the velocity field between the upper and lower disk samples is valid,
because the disk scale height is about 300 pc ($\sim 3^{\circ}$ at 6 kpc);
the off-plane sample ($|b|>3^{\circ}$) does not contain many distant disk objects 
above the disk scale height.

We have made a two-dimensional Kolmogorov-Smirnov (K-S) test 
\citep{fas87} for the upper and lower disk samples [for algorithm for the two-sample test, 
see section 14.7 of \citet{pre92}].
We found that the null hypothesis of distribution in the longitude-velocity plane 
for the off-plane sample being the same as distribution of the on-plane sample
has a probability of less than 0.1 \%. Therefore, the conjecture that the
off- and on-plane samples have different velocity distributions is verified 
with a 99.9 \% confidence level.
The largest difference in fraction between the two samples is 0.386; this
occurs in the upper-left quadrant at the point $(l,V_{lsr})=$(13.7$^{\circ}$, 30.9 km s$^{-1}$),
which is indicated by a cross in the upper panel of figure 5. Most of the objects on
the right side of this point ($l<14^{\circ}$) supposedly belong to the
Galactic bulge, where the velocity field is different from that of the disk objects.
Most of the objects in the left portion ($l>14$ $^{\circ}$) are expected to be disk sources.
 
The lack of high velocity objects in the off-plane sample appears more strongly 
for the subsample of the $b<-3^{\circ}$, $l>14^{\circ}$ objects. 
The lower panel of figure 5 shows a comparison between
the velocity distributions of objects with $b<-3^{\circ}$ and $b>3^{\circ}$. 
Above $l=14^{\circ}$, only one object in the $b<-3^{\circ}$ \ subsample 
has $V_{lsr}>50$ km s$^{-1}$, while there are 6 objects in the $b>3^{\circ}$ subsample
with such velocities (though the two-dimensional K-S test for these sets 
does not give any statistical significance due to the small numbers).   

This asymmetry with respect to the Galactic latitude for the off-plane objects can be
attributed to the warp of the Galactic disk. Because the warped disk plane is shifted toward
positive $b$ in $0\lesssim l \lesssim 180^{\circ}$, a shift of the warped disk plane
of less than a degree would considerably influence the cut off of the distant objects 
in off-plane subsamples.
Though, in general, the disk warp is considered to appear more strongly for the outer disk samples
\citep{djo89}, it also appears in inner disk samples [for example, see the case for MSX point sources
by \citet{vig05}]. 
Figure 6 shows the distribution of sources in the Galactic coordinates and the regression curves
for several subsamples. Apparently, in order to obtain the better regression-curve fit, 
the sample involving distant objects with higher $|b|$ is preferable, 
so that the upper disk sample in this paper is quite useful to limit the 
warp parameters. Table 5 lists the bet-fit parameters
for three different samples: the on-plane sample ($|b|<3^{\circ}$),
the whole sample ($|b|<5^{\circ}$), and high-velocity ($V_{\rm LSR} >50$ km s$^{-1}$) sample.
Both the sense of the inclination ($b_0\sim 2^{\circ}$) and the shift in the azimuthal angle ($\phi_0\sim 20$ $^{\circ}$) 
of the warp plane match to those obtained by \citet{vig05} analyzing the inner-disk MSX PSC sources.
 
An alternative explanation of this asymmetric star distribution 
(especially for the nearby star concentration around $l\sim 23^{\circ}$ and $b\sim -4^{\circ}$)  
could be accretion of a dwarf galaxy debris by the Galaxy. However, the high stellar density
of this asymmetric star group in the inner disk virtually excludes such a possibility, though in general 
it may be hard to distinguish such accretion process from the effect of warp in the outer disk
at low star density [for example, arguments in \cite{mom06} for the case of CMa over-density]. 

\subsection{A group of stars with unusual velocities: a star stream}
We noticed a concentration of three stars with $V_{LSR}<-50$ km s$^{-1}$ in a small area 
in figure 3 (indicated by filled triangles at lower left);  
$J$19222258$-$1418050 ($-$74.6 km s$^{-1}$), $J$19291840$-$1916199 ($-$68.8 km s$^{-1}$)
$J$19310440$-$1640135 ($-$69.4 km s$^{-1}$).
They are concentrated within a few degrees at $(l,b)\sim (21.7^{\circ}, -15.4^{\circ})$ .
This is quite unusual, because the Galactic rotation at this longitude should give a positive $V_{\rm LSR}$
for most of the objects (except for the stars located very far beyond 15 kpc).
These objects are medium bright sources ($F_{C}=8$--24 Jy and
$K=4.2$--5.5), so that their distances are estimated to be less than 7 kpc.
We also noticed that some disk maser sources ($|b|<3^{\circ}$) occasionally 
exhibit similarly large negative velocities in the area $15^{\circ}<l<30^{\circ}$;
for example, see the SiO and OH $l$--$v$ diagrams (figure 4) given by \citet{deg04b}.  
In table 6, we list infrared properties of these objects with radial velocity less than $-50$ km s$^{-1}$
located in  $19^{\circ}<l<32^{\circ}$, giving the IRAS name, the alternative name,
Galactic coordinates, $V_{\rm LSR}$, 2MASS $K$ magnitude, $H-K$, and corrected $K$ magnitude $K_c$, 
IRAS 12 $\mu$m flux density $F_{12}$ and the IRAS color $C_{12} [=log(F_{25}/F_{12})]$. 

The positions of these objects are plotted 
in Galactic coordinates in figure 7. But the CO $l$--$v$ diagram does not show any feature
associated with these velocities in this longitude range \citep{dam01,kaw99}, although there is
a distant spiral arm feature spreading at higher velocities 
[from ($\sim 20^{\circ}$, 0 km s$^{-1}$) to ($\sim 40^{\circ}$, $-50$ km s$^{-1}$]
(due to the 15 kpc arm). We also do not find any negative-velocity SiO maser sources 
with $V_{\rm LSR}<-50$ km s$^{-1}$
up to the longitude $l=50^{\circ}$, where distant objects obeying the Galactic rotation start to show
negative $V_{\rm LSR}$ (for example see figure 8 of \cite{nak03}).
Note that the radial velocities of these stars, if motion of the Local Standard of Rest (220 km s$^{-1}$) 
about the Galactic center is subtracted (i.e., velocities in the GSR), are
quite small ($\sim 20$ km s$^{-1}$), suggesting that these stars are moving  
perpendicularly to the line of sight with a high speed. 

The origin of this negative velocity group is not clear. We have several 
conjectures. One is that they are a part of the radial outward stellar motion, which is produced
by the dynamical effect of the bulge bar. According to an analysis of Hipparcos data made by
\citet{fea00}, a stream of short-period Mira stars is observed at negative velocities 
($\sim -75$ km s$^{-1}$) toward the bulge 
and some of these extend beyond the solar circle.   
The negative-velocity streaming of these stars spreads in a large scale toward the bulge.
The feature we observe, especially around $l\sim 30^{\circ}$ might be a part of
this large-scale outward star flow; the concentration around $(l,b)\sim (22^{\circ}, -15^{\circ})$ 
may be a stochastic effect.

An alternative conjecture is that the velocity structure 
is a manifestation of a star stream crossing in the Galactic disk (for example, see \cite{nav04}). 
The broken line in figure 7 
indicates a linear least-squares fit to the positions of the negative velocity objects.
We found that this regression line passes within 0.1$^{\circ}$ of NGC 6712. 
This globular cluster has $V_{\rm LSR}=-92$ km s$^{-1}$ \citep{har96},
that is close to the average velocity
of these objects ($-69.3$ km s$^{-1}$) if the uncertainty of optical radial velocity measurements
of globular-cluster members (of about 20 km s$^{-1}$) is factored in \citep{mat05}. 
The other 2 nearby globular clusters 
have somewhat similar radial velocities ($V_{\rm LSR}=-53.9$ km s$^{-1}$ 
and $-44.9$ km s$^{-1}$ for Pal 11 and NGC 6749, respectively)
except for NGC 6760 ($V_{\rm LSR}=10.9$ km s$^{-1}$). It is possible that these globular clusters 
are fossil remnants of  the same dwarf galaxy plunging into 
the dense inner disk of the Galaxy. In fact, \citet{dem99} found evidence of tidal disruption of NGC 6712. 
\citet{pal01} found about one hundred blue stragglers in the same globular cluster. 
In the core of globular clusters, it is possible to generate 
the antecedents of AGB stars by main-sequence mergers \citep{mat05,and06},
 and later they can escape out from the original globular cluster by tidal interactions.  
Because primordial low-mass  metal-poor stars in aged globular clusters ($M <1 M_{\odot}$) cannot be evolved 
to SiO maser stars with large mass-loss rates at the AGB phase, only larger mass stars
such as merger antecedents can be sources of maser stars in globular clusters or dwarf galaxies.
Although, at present, we do not have any other strong evidence of existence of the dwarf galaxy or
it's tidal tail in this direction,  the presence of a star stream remains to be an interesting speculation.


\section{Summary}

We surveyed 277 2MASS/MSX or IRAS sources off the Galactic plane
and obtained more than one hundred new detections in the SiO maser lines. 
Two of these SiO sources are associated with very faint infrared objects, and are likely
to be associated with the Sgr or CMa dwarf galaxies. We also found a clear signature of a
 warp of the Galactic disk in the off-plane ($|b|>3^{\circ}$) sample.
Furthermore, we found an interesting group of O-rich AGB stars 
at $(l, b) \sim (22^{\circ}, -15^{\circ})$, which do not follow the Galactic rotation.
This may suggest the presence of a stream of stars penetrating the Galactic disk,
whose origin is not clear.
 
\

Authors thank Dr. B. M. Lewis for the useful comments and kind advices on English.
One of authors (A.W.) aknowledges National Astronomical Observatory of Japan for approving him 
a short-term visiting fellowship in 2003. 
This research made use of the SIMBAD and VizieR databases operated at CDS, 
Strasbourg, France, as well as use of data products from 
Two Micron All Sky Survey, which is a joint
project of the University of Massachusetts and Infrared Processing 
and Analysis Center/California Institute of Technology, 
funded by the National Aeronautics and Space Administration and
National Science foundation, and from the Midcourse Space 
Experiment at NASA/ IPAC Infrared Science Archive, which is operated by the 
Jet Propulsion Laboratory, California Institute of Technology, 
under contract with the National Aeronautics and Space Administration.   

\section*{Appendix. Individual Objects}
The SiO maser spectra for the newly detected objects are shown in Figure 8a--8g. 
We discuss here interesting objects individually.

\begin{itemize}
\item
$J$06353168$-$0109256 (IRAS 06329$-$0106): This object has a radial velocity
100.4 km s$^{-1}$ at $l=212^{\circ}$, which is considerably large if compared with the Galactic
rotational velocity (see the lower panel of figure 3). The SiO masers were detected first by \citet{jia96}.
This object is a bright infrared source ($K=3.7$ and $F_{12}=18.8$ Jy) with IRAS LRS class 29 (silicate emission ; \cite{oln86}),
suggesting a distance of a few kpc. It is unlikely 
that this is associated to CMa dwarf or its tail.

\item
$J$17572314$-$1802456 (IRAS 17544$-$1802):  This object is denoted as
LDN 292, a dark Nebula, in the SIMBAD database. However, the detection
of SiO masers secures this object as an AGB star, which is consistent
with its IRAS LRS class of 29 (silicate emission). The 2MASS images
exhibit a faint nebulosity of size of about 30$''$ toward this bright (K=3.546) object. 

\item
$J$17001324$-$1037003 (IRAS 16574$-$1032 =V2207 Oph =IRC $-10355$): This is a strong IRAS source
with $F_{12}=89.6$ Jy and LRS class of 28 (silicate emission). 
Previous searches for OH 1612 MHz and H$_2$O 22 GHz emission
were negative \citep{tel91,lew97}. \citet{nym92} included this object in the list 
of tentative detections by the CO $J=1$--0 line, but the radial velocity was not given.
NIR light variations have been detected \citep{loc85}. 
The detection of strong SiO masers at $V_{lsr}=-42.0$ km s$^{-1}$ confirms that
this is a mass-losing AGB star.

\item
$J$17335931$-$2659216 (IRAS 17308$-$2657): This object has a large negative radial velocity of $-261$ km s$^{-1}$.
Its position [$(l,b)=(0.3^{\circ}, +3.2^{\circ})$] and infrared flux densities ($F_C=7.7$ Jy and $K=7.9$) indicate
that this is a star in the Galactic bulge. Large negative radial velocities are common
for bulge stars in the $x_1$ orbit family \citep{fuj06}.

\item
$J$17415458$-$2047055 (IRAS 17389$-$2045): 
This is a strong infrared source with $F_{12}=66.0$ Jy and LRS class of 28 (silicate emission).
OH 1612 MHz and 1665/67 MHz lines have been detected (\cite{les92,dav93})
at $V_{lsr}=13.8$ and 47.5 km s$^{-1}$, the mean velocity being consistent 
with the SiO detection at $V_{lsr}=31.3$ km s$^{-1}$.
This object was included in the relatively small number of the AGB stars
sampled by IRTS; from infrared spectra between 1--5 $\mu$m,
the distance was estimated to be about 1 kpc (\cite{leb01}).

\item
$J$18085198$-$2616210 (IRAS 18057$-$2616):
This is a strong infrared source with $F_{12}$=59 Jy with LRS class of 26 (silicate emission).
An OH 1612 MHz maser was detected at $V_{\rm lsr}=-7.7$ and 29.7 km s$^{-1}$ (\cite{tel91}),
the mean velocity being consistent with SiO radial velocity of 11.1 km s$^{-1}$. 

\item
$J$23584412$-$3926599 (=RR Phe) : 
Though it is a well-known optical/NIR variable star
with a period of 427 days (\cite{jur93}) located 
close to the South Galactic pole ($b=-73.5^{\circ}$),
it was not recorded in the IRAS catalog because of its position 
in the IRAS gap and had been neglected. 
The MSX out-of-plane survey found this object as a strong MIR source with $F_{C}=$ 289 Jy.
We detected relatively strong SiO maser emission at $V_{\rm lsr}$=18 km s$^{-1}$.
No previous OH or H$_2$O maser observation was found for this star.

\end{itemize}

\tabcolsep 2pt
\begin{longtable}{l@{\hspace{-1mm}}rccc@{\hspace{2mm}}rccc@{\hspace{2mm}}rr}
\caption{SiO detections.}
\hline \hline
%
    & \multicolumn{4}{c}{SiO $J=1$--0 $v=1$} & \multicolumn{4}{c}{SiO $J=1$--0 $v=2$} & & \\
	\cline{2-5} \cline{6-9}
2MASS name & $V_{\rm peak}$ & $T_{\rm peak}$ & Flux & RMS & $V_{\rm peak}$ & $T_{\rm peak}$ & Flux & RMS & Obs. date & $V_{\rm ave}$\\     & {\tiny (km s$^{-1}$)} & {\tiny (K)} & {\tiny (K km s$^{-1}$)} & {\tiny (K)} 
	 & {\tiny (km s$^{-1}$)} & {\tiny (K)} & {\tiny (K km s$^{-1}$)} & {\tiny (K)} & {\tiny (yymmdd.d)} & {\tiny (km s$^{-1}$)} \\
\hline
\endfirsthead
\hline \hline
    & \multicolumn{4}{c}{SiO $J=1$--0 $v=1$} & \multicolumn{4}{c}{SiO $J=1$--0 $v=2$} & & \\
	\cline{2-5} \cline{6-9}
2MASS name & $V_{\rm peak}$ & $T_{\rm peak}$ & Flux & RMS & $V_{\rm peak}$ & $T_{\rm peak}$ & Flux & RMS & Obs. date & $V_{\rm ave}$\\
     & {\tiny (km s$^{-1}$)} & {\tiny (K)} & {\tiny (K km s$^{-1}$)} & {\tiny (K)} 
	 & {\tiny (km s$^{-1}$)} & {\tiny (K)} & {\tiny (K km s$^{-1}$)} & {\tiny (K)} & {\tiny (yymmdd.d)} & {\tiny (km s$^{-1}$)} \\
\hline
\endhead
\hline
\endfoot
\hline
\multicolumn{11}{l}{$^{\dag}$ detected previously by \citet{jia96}. } \\
\multicolumn{11}{l}{$^{\ddag}$ detected previously by \citet{nym98}. } \\
\multicolumn{11}{l}{$^{\S}$ the SiO $J=2$--1 $v=1$ emission detected previously by \citet{deg01}. } \\
\multicolumn{11}{l}{$^{\flat}$ detected previously by \citet{ita01}.} \\
\endlastfoot
$J$02160009$-$2031108 &   4.1 & 0.362 &  0.722 & 0.072 &    $-$1.3 & 0.950 &  1.515 & 0.096 & 030516.5 &    1.4  \\  
$J$03055291$+$5420538 & $-$28.6 & 0.249 & 0.387 & 0.043 & $-$28.0 & 0.367 & 0.424 & 0.049 & 990516.5 & $-$28.3 \\ 
$J$05475868$-$3305109 & 110.8 & 0.114 &  0.471 & 0.030 &    ---    &  ---  &  ---   & 0.039 & 040226.8 &  110.8  \\
$J$06215148$-$0350416$^{\dag}$ & $-$7.3 & 1.541 & 5.152 & 0.053 & $-$7.3 & 1.456 & 4.861 & 0.107 & 000115.9 & $-$7.3 \\ 
$J$06353168$-$0109256$^{\dag}$ & 100.1 & 1.103 & 2.336 & 0.094 & 100.7 & 0.932 & 2.070 & 0.101 & 000407.6 & 100.4 \\ 
$J$06421530$-$0939442$^{\dag}$ & 26.4 & 3.954 & 10.679 & 0.100 & 26.4 & 4.503 & 10.071 & 0.123 & 000407.7 & 26.4 \\ 
$J$06425082$-$0141061$^{\dag}$ & 17.8 & 0.297 & 0.717 & 0.060 & 18.3 & 0.543 & 1.128 & 0.068 & 000116.9 & 18.0 \\ 
$J$07072736$-$2538440 & 84.9 & 1.006 & 5.406 & 0.076 & 85.1 & 0.835 & 4.324 & 0.079 & 000116.9 & 85.0 \\
\end{longtable}

\begin{longtable}{lrrc}
  \caption{Negative results.} 
\hline  \hline
name & RMS($v=1$) & RMS($v=2$) & Obsdate \\
     & (K)     &  (K)     & (yymmdd.d) \\
\hline
\endfirsthead
\hline \hline
name & RMS($v=1$) & RMS($v=2$) & Obsdate \\
     & (K)     &  (K)     & (yymmdd.d) \\
\hline
\endhead
\hline
\endfoot
\hline
\endlastfoot
$J$01015839$+$5759482 & 0.053 & 0.059 & 990515.5 \\ 
$J$02005614$+$0945356 & 0.037 & 0.027 & 030516.4 \\ 
$J$02151393$+$5613030 & 0.069 & 0.074 & 990515.5 \\ 
$J$02300585$+$4007038 & 0.039 & 0.030 & 030516.3 \\ 
$J$02311288$+$6340198 & 0.102 & 0.213 & 990515.6 \\ 
$J$03133515$+$5948036 & 0.049 & 0.046 & 990517.5 \\ 
$J$03230583$+$5652435 & 0.043 & 0.040 & 990517.6 \\ 
$J$03275901$+$6044553 & 0.038 & 0.031 & 990516.6 \\
\end{longtable}

\tabcolsep 1pt
\begin{longtable}{rrrccccccccc}
\caption{Infrared properties of the observed sources} 
\hline
2MASS name  & $K$ & $J-H$ & $H-K$ & Bflag & MSX6C name & $\Delta r_{M}$ & $F_C$ & $C_{AC}$ & $C_{CE}$ & IRAS name & $\Delta r_{I}$ \\
            & (mag) & (mag) & (mag) &  &           & ($''$)  & (Jy)  &        &          &            & ($''$)  \\
\hline 
\endfirsthead
\hline
2MASS name  & $K$ & $J-H$ & $H-K$ & Bflag & MSX6C name & $\Delta r_{M}$ & $F_C$ & $C_{AC}$ & $C_{CE}$ & IRAS name & $\Delta r_{I}$ \\
            & (mag) & (mag) & (mag) &  &           & ($''$)  & (Jy)  &        &          &            & ($''$)  \\
\hline
\endhead
\hline

\endfoot
\hline
\multicolumn{12}{l}{$^{\dag}$ indicates IRAS 12 $\mu$m flux density in column $F_C$ and }\\
\multicolumn{12}{l}{logrithmic ratio of IRAS 25 to 12 $\mu$m flux-densities in column $C_{CE}$.}\\
\multicolumn{12}{l}{$^{\ddag}$ no 2MASS object, but the IRAS position is shown.}\\
\endlastfoot
$J$01015839$+$5759482 & 4.123 & 0.864 & 0.448 & 111 & (124.332$-$04.846) &  & 4.3$^{\dag}$ &  & $-$0.228$^{\dag}$ & 00589$+$5743 & 5.1 \\ 
$J$02005614$+$0945356 & 7.170 & 1.741 & 1.355 & 111 & (149.845$-$49.446) &  &  2.3$^{\dag}$ &  & $-$0.50$^{\dag}$ & 01582$+$0931 & 1.4 \\
$J$02151393$+$5613030 & 6.014 & 1.747 & 1.111 & 111 & (134.439$-$04.771) &  & 7.4$^{\dag}$ &  & $-$0.040$^{\dag}$ & 02117$+$5559 & 2.5 \\ 
$J$02160009$-$2031108 & 2.613 & 0.864 & 0.502 & 111 & (198.593$-$69.596) &  & 26.7$^{\dag}$ &  & $-$0.20$^{\dag}$ & 02136$-$2045 & 7.1 \\
$J$02300585$+$4007038 &11.701 & 3.039 & 2.745 &  11 & (142.651$-$18.946) &  &  1.0$^{\dag}$ &  & $-$0.16$^{\dag}$ & 02269$+$3953 & 0.6 \\
$J$02311288$+$6340198 & 6.272 & 2.452 & 1.896 & 111 & G133.7260$+$02.9110 & 1.5 & 6.13 & $-$0.069 & $-$0.346 & 02272$+$6327 & 3.1 \\ 
$J$03055291$+$5420538 & 4.950 & 1.343 & 0.773 & 111 & G141.7917$-$03.5248 & 1.3 & 7.60 & $-$0.003 & $-$0.099 & 03022$+$5409 & 5.7 \\ 
$J$03133515$+$5948036 & 6.217 & 2.739 & 2.119 & 111 & G139.9388$+$01.7086 & 1.1 & 12.25 & 0.003 & $-$0.198 & 03096$+$5936 & 5.0 \\
\end{longtable}

\tabcolsep 2pt
\begin{longtable}{lcrcccccccc}
\caption{Properties of the observed sources for dwarf galaxy member candidates.} 
\hline
2MASS name  & Group & $V_{GSR}^{SiO}$ & $\Lambda_{\odot}$ & $\beta_{\odot}$  & $K$   & $H-K$ & $A_K$ & $D^{\S}$ & $F_C$ & IRAS name \\
            &       & (km s$^{-1}$)     & (deg)   &  (deg)   & (mag) & (mag) &  (mag) & (kpc) & (Jy) &          \\
\hline 
\endfirsthead
\hline
2MASS name  & Group & $V_{GSR}^{SiO}$ & $\Lambda_{\odot}$ & $\beta_{\odot}$  & $K$   & $H-K$ & $A_K$ & $D^{\S}$ & $F_C$ & IRAS name \\
            &       & (km s$^{-1}$)     & (deg)   &  (deg)   & (mag) & (mag) &  (mag) & (kpc) & (Jy) &          \\
\hline
\endhead
\hline

\endfoot
\hline
\multicolumn{10}{l}{$^{\S}$ P-M relation, $M_K=6.43$ at 8 kpc for P=450 days, was used.}\\
\multicolumn{10}{l}{$^{*}$ from \citet{mau05}.}\\
\endlastfoot
$J$02005614$+$0945356 & Sgr &         & 109.1 &$-$10.1 & 7.170  & 1.355 & 0.025 &    7 &  2.3  &  01582$+$0931   \\
$J$02300585$+$4007038 & Sgr &         & 133.0 &$-$32.0 &11.701  & 2.745 & 0.021 &   20 &  1.0  &  02269$+$3953   \\
$J$03482812$+$1657032 & Sgr &         & 135.6 & $-$3.3 &11.097  & 0.806 & 0.116 &$>30$ &       &                 \\
$J$03552337$+$1133437 & Sgr &         & 134.6 &    2.2 &11.526  & 1.004 & 0.088 &$>30$ &       &                 \\
$J$03595596$+$0919044 & Sgr &         & 134.5 &    4.7 &10.114  & 1.115 & 0.101 &   29 &       &                 \\
$J$05475868$-$3305109 & CMa & $-56.0$ & 138.3 &   54.2 & 6.152  & 0.656 & 0.012 &    7 & 2.0   &  05461$-$3306   \\
$J$06513922$-$2123303 & CMa &         & 166.9 &   49.1 & 6.834  & 1.560 & 0.108 &    5 & 3.1   &  06495$-$2119   \\
$J$06552346$-$2122258 & CMa &         & 168.2 &   49.3 & 6.576  & 1.454 & 0.119 &    5 & 3.4   &  06532$-$2118   \\
\end{longtable}

\clearpage
\begin{table*}
\begin{center}
  \caption{The values of the best parameters of the warp from SiO maser data.}\label{tab:table3}
  \begin{tabular}{lcrrr}
  \hline\hline
Sample   & Data points  & $b_0$ & $\phi_0$  & $R$  \\
         &              & (deg) & (deg) &  \\
\hline   
Disk only ($|b|<3^{\circ}$) &  79 & 0.50 ($\pm 1.16$) & 25.6 ($\pm 17.5$) & 0.049 \\
Whole ($|b|<5^{\circ}$)     & 108 & 3.09 ($\pm 1.62$) & 28.4  ($\pm 4.0$) & 0.182 \\
Velocity limited ($|b|<5^{\circ}$ with $V_{lsr}>50$ km s$^{-1}$)  & 48 &  2.23 ($\pm 2.27$) &  19.8 ($\pm 10.1$)  &0.141 \\
\hline  
\end{tabular}  
\end{center}  
\end{table*}

\tabcolsep 2pt
\begin{longtable}{lccccccccc}
\caption{Properties of maser sources with $V_{\rm LSR}<-50$ km s$^{-1}$.} 
\hline
IRAS name & Other name & $l$ & $b$ & $V_{\rm LSR}$ & $K$ & $H-K$  & $K_c$   & $F_{12}$ & $C_{12}$ \\
          &            & (deg)& (deg) & km s$^{-1}$ & (mag) & (mag) &  (mag) & (Jy) & \\
\hline 
\endfirsthead
\hline
IRAS name & Other name & $l$ & $b$ & $V_{\rm LSR}$ & $K$ & $H-K$  & $K_c$   & $F_{12}$ & $C_{12}$ \\
          &            & (deg)& (deg) & km s$^{-1}$ & (mag) & (mag) &  (mag) & (Jy) &  \\
\hline
\endhead
\hline

\endfoot
\hline
\multicolumn{10}{l}{$^{*}$ \cite{sev01}}\\
\multicolumn{10}{l}{$^{\dagger}$ \cite{ede88}}\\
\multicolumn{10}{l}{$^{\ddagger}$ \cite{tel91}}\\
\multicolumn{10}{l}{$^{\P}$ Present work.}\\
\multicolumn{10}{l}{$^{\flat}$ not identified in 2MASS database, but seen on the Spitzer GLIMPSE image; see \cite{deg07}.}\\
\multicolumn{10}{l}{$^{\S}$ Apparent double sources; flux densities not reliable \citep{deg99}.}\\
\endlastfoot
18307$+$0102 & OH031.8+04.5     & 31.817 &   4.537 & $-$62.8$^{\dagger}$ & 6.902 & 1.561 & 5.374 & 6.4 & 0.111 \\
18325$-$0721 & OH024.581+00.224 & 24.581 &   0.224 & $-$72.6$^{*}$ & ---$^{\flat}$ & --- & --- & 5.3 & 0.340 \\
18387$-$0951 &                  & 23.078 &  $-$2.274 & $-62.8^{\ddagger}$ & 12.215 & 1.075 & 11.387 & 7.0 & 0.299 \\
18393$-$1020 & OH022.707$-$02.631 & 22.707 &  $-$2.631 & $-$76.1$^{*}$ & 6.934 & 1.400 & 5.638 & 5.9 & $-$0.070 \\
18399$-$0149 & OH030.336+01.171 & 30.336 &   1.171 & $-$64.7$^{*}$ & 4.607 & 0.899 & 4.032 & 12.1 & $-$0.134 \\
18465$-$0717 & OH026.246$-$02.814 & 26.246 &  $-$2.814 & $-$65.8$^{*}$ & 7.886 & 2.876 & 4.465 & $16.4^{\S}$ & $0.010^{\S}$ \\
19195$-$1423 &                  & 23.448 & $-$13.220 & $-$75.3$^{\ddagger\P}$ & 5.598 & 1.400 & 4.302 & 8.8 & 0.170 \\
19263$-$1922 &                  & 19.466 & $-$16.779 & $-$68.8$^{\P}$ & 4.220 & 0.570 & 4.119 & 11.9 & $-$0.190 \\
\end{longtable}


\begin{figure*}
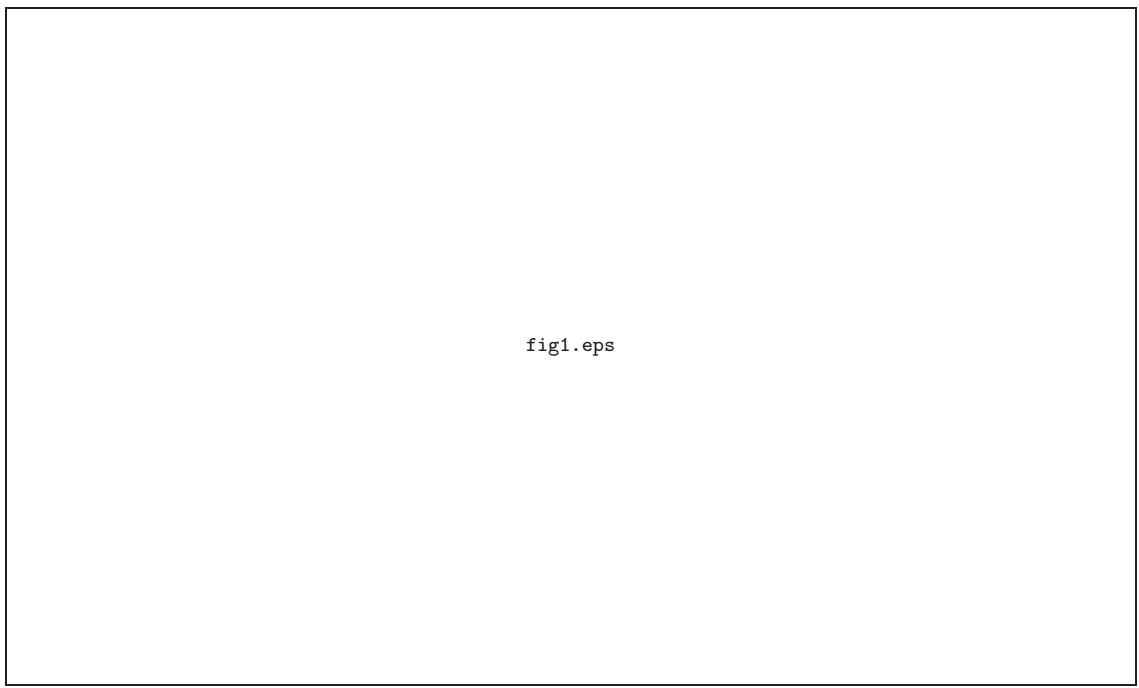

  \begin{center}
    \FigureFile(150mm,90mm){fig1.eps}
  \end{center}
  \caption{Distribution of the observed sources in the Galactic coordinates in the Hammer-Aitoff's projection.
  The dotted curve indicates the orbital plane of SagDEG.
}\label{fig: l-b}
\end{figure*}


\begin{figure*}
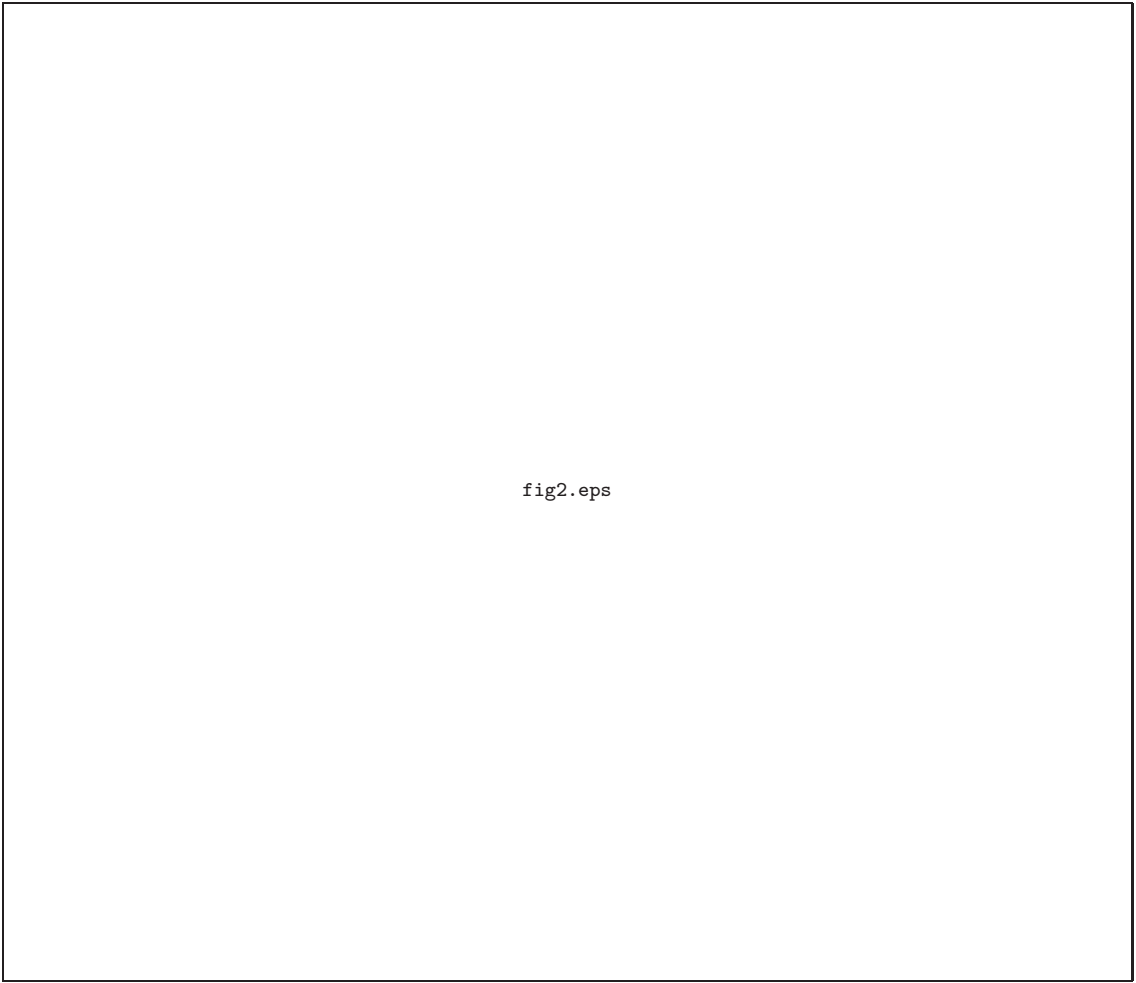

  \begin{center}
    \FigureFile(150mm,130mm){fig2.eps}
  \end{center}
  \caption{Plots of $K$ versus $H-K$ and $F_C$ versus $C_{CE}$.
  Filled and unfilled circles indicate SiO detection and nondetection.
}\label{fig: flux-color}
\end{figure*}


\begin{figure*}
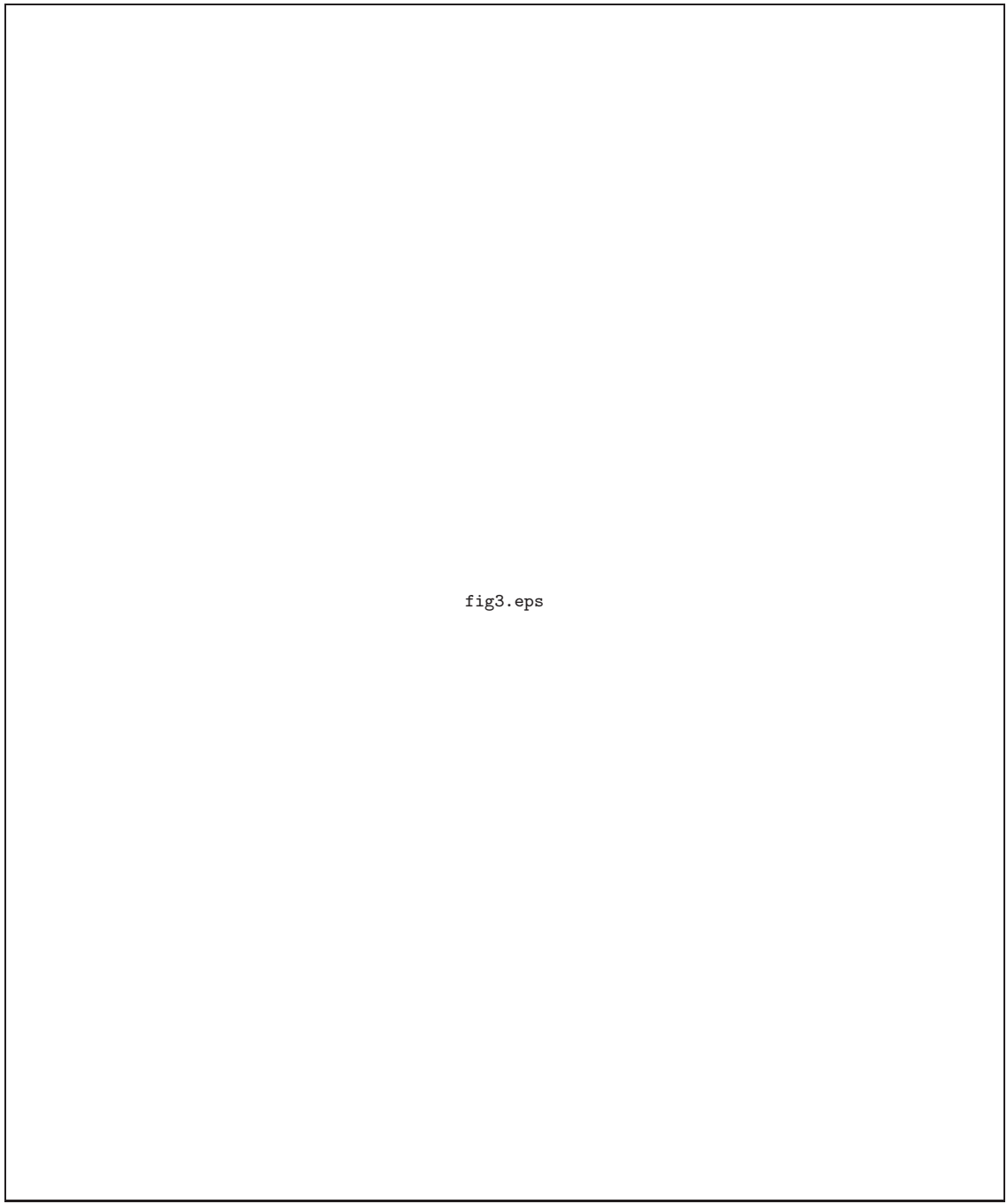

  \begin{center}
    \FigureFile(150mm,180mm){fig3.eps}
  \end{center}
  \caption{Velocity distribution of the observed sources in the ranges $-40^{\circ}<l<50^{\circ}$
  (top) and $180^{\circ}<l<270^{\circ}$ (bottom). The solid and broken curves indicate the velocities
  expected from the Galactic rotation curve, which is assumed to be flat at 220 km s$^{-1}$
  or linear within 1 kpc from the Galactic center.
  One object with $V_{\rm lsr}=-261.5$ km s$^{-1}$ ($J17335931-2659216$), is out of this figure.
  The star J05475868 is a candidate for the dwarf galaxy association, and the stars J06353168 and J08144298 
  are likely disk objects (see section 3.1 and appendix).
}\label{fig: velocity1}
\end{figure*}


\begin{figure*}
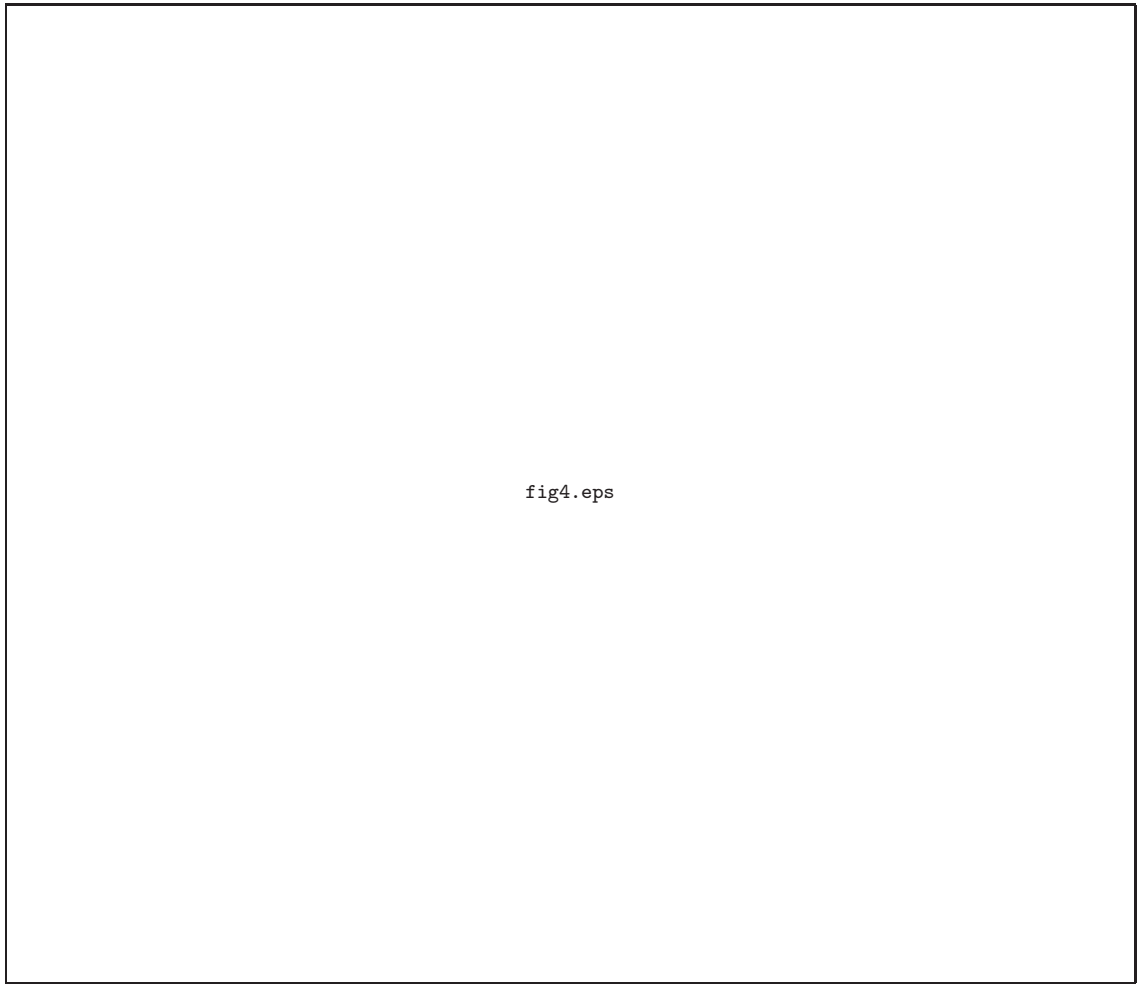

  \begin{center}
    \FigureFile(150mm,130mm){fig4.eps}
  \end{center}
  \caption{Velocity distribution of the SiO maser sources in the Sgr coordinates. Blue diamonds indicate the SiO maser sources
  detected in this paper, and the large filled red circle indicates the candidate 
  for SagDEG tidal-tail associations, $J$1923554$-$1302029. Black squares indicate the M giants sampled by \citet{maj04} and
  the thin broad bands are result of simulation of the tidal tail of SagDEG, which were taken from \citet{law05}.  
}\label{fig: Sgrlamda}
\end{figure*}


\begin{figure*}
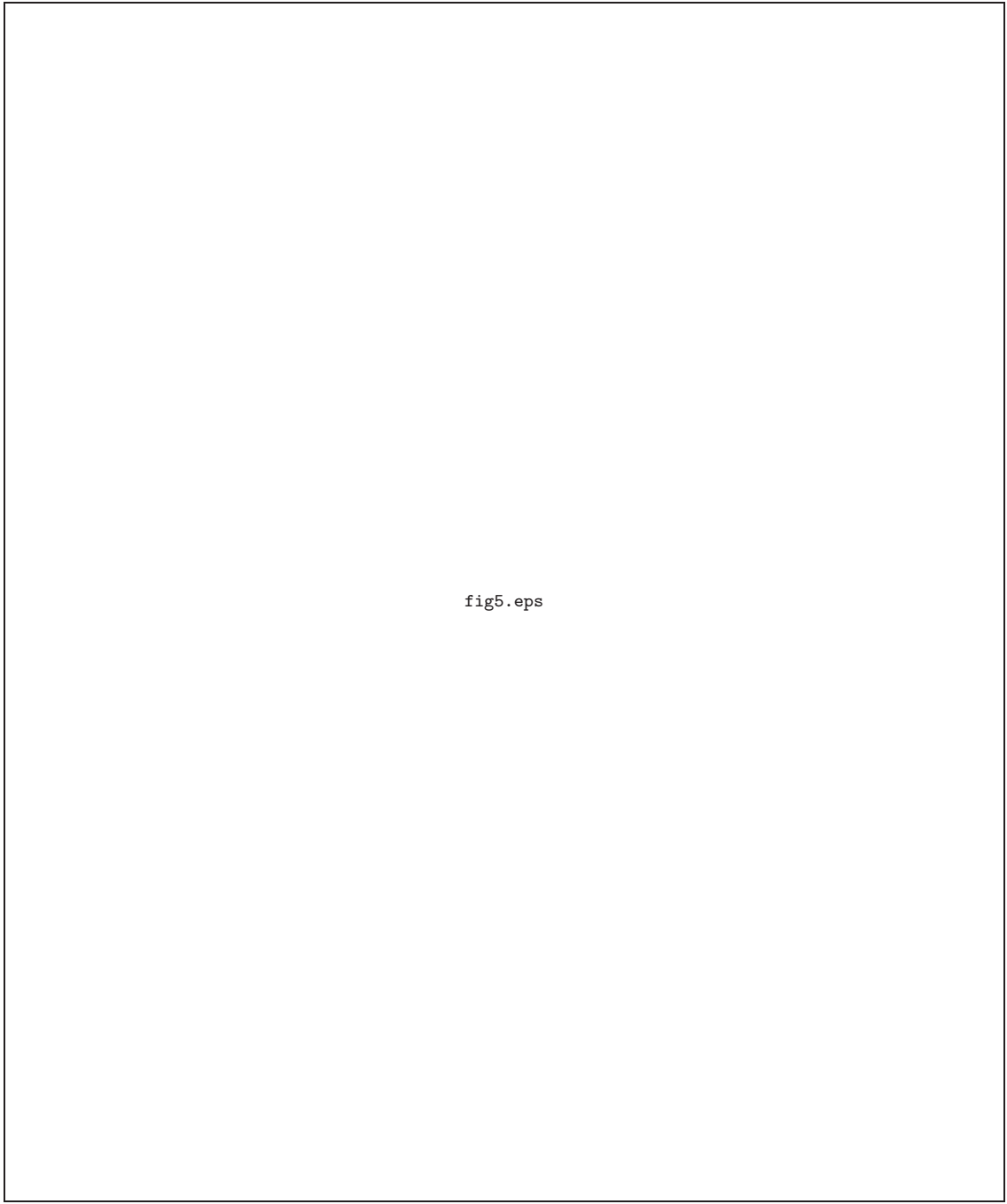

  \begin{center}
    \FigureFile(150mm,180mm){fig5.eps}
  \end{center}
  \caption{Comparison of velocity distributions for objects above and below $|b|=3^{\circ}$ (top),
  and another comparison for upper disk objects with $b>3^{\circ}$ and $b<-3^{\circ}$ (bottom).
  All the objects plotted have $|b|< 6^{\circ}$.    
  The cross in the top panel indicates the position of the most significant
  difference in fraction between two samples from a 2-$d$ Kolmogorov-Smirnov test.
}\label{fig: velocity2}
\end{figure*}

\begin{figure*}
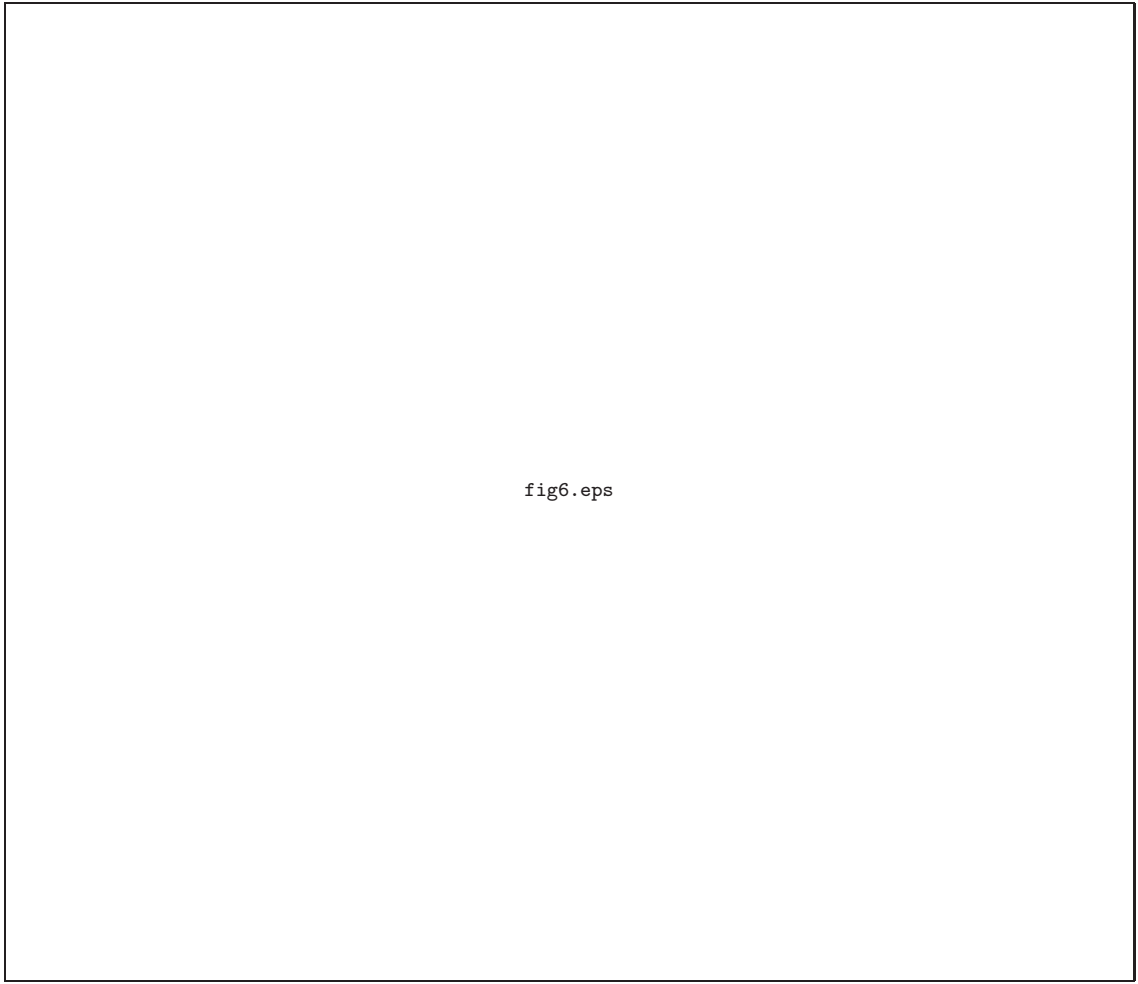

  \begin{center}
    \FigureFile(150mm,130mm){fig6.eps}
  \end{center}
  \caption{Source distribution in Galactic coordinates. The small and large open circles
  indicate the disk sample ($|b|<3^{\circ}$) and whole sample ($|b|<5^{\circ}$).
  Filled circles indicates a subsample of distant objects ($V_{lsr}>50$ km s$^{-1}$).
  There is a concentration of 4 sources within 1 degree at (26.3$^{\circ}$, 3.3$^{\circ}$), but this seems a chance coincidence,
  because their radial velocities are very diverse from $-15$ to 86 km s$^{-1}$. The thin ellipse
  at the lower right part indicates a concentration of relatively nearby stars asymmetric with respect to the Galactic plane.
  The best-fit linear line for each sample [minimizing square sum of residue 
  $y_i=b_i -b_0\times sin (l_i- \phi_0), \  \  \;  \;  (i=1, ... , n)$]
  is shown. 
}\label{fig: velocity3}
\end{figure*}

\begin{figure*}
  \begin{center}
    \FigureFile(150mm,130mm){fig7.eps}
  \end{center}
  \caption{Distribution of SiO/OH maser sources with $V_{\rm LSR}<-50$ km s$^{-1}$ (filled circles). 
  Positions of the nearby Globular clusters are indicated by open squares; 
  data are taken from \citet{har96} except GLIMPSE C01 \citep{kob05}. The broken line indicates
  the least squares fit to the Galactic coordinates of maser sources;
  $l= 0.39 \ (\pm 0.11) \times b + 27.0^{\circ} \ (\pm 1.1^{\circ}) $, 
  where the values between parentheses are standard errors of the coefficients.
}\label{fig: lb-cluster}
\end{figure*}

\newpage
\setcounter{figure}{7}
\begin{figure*}
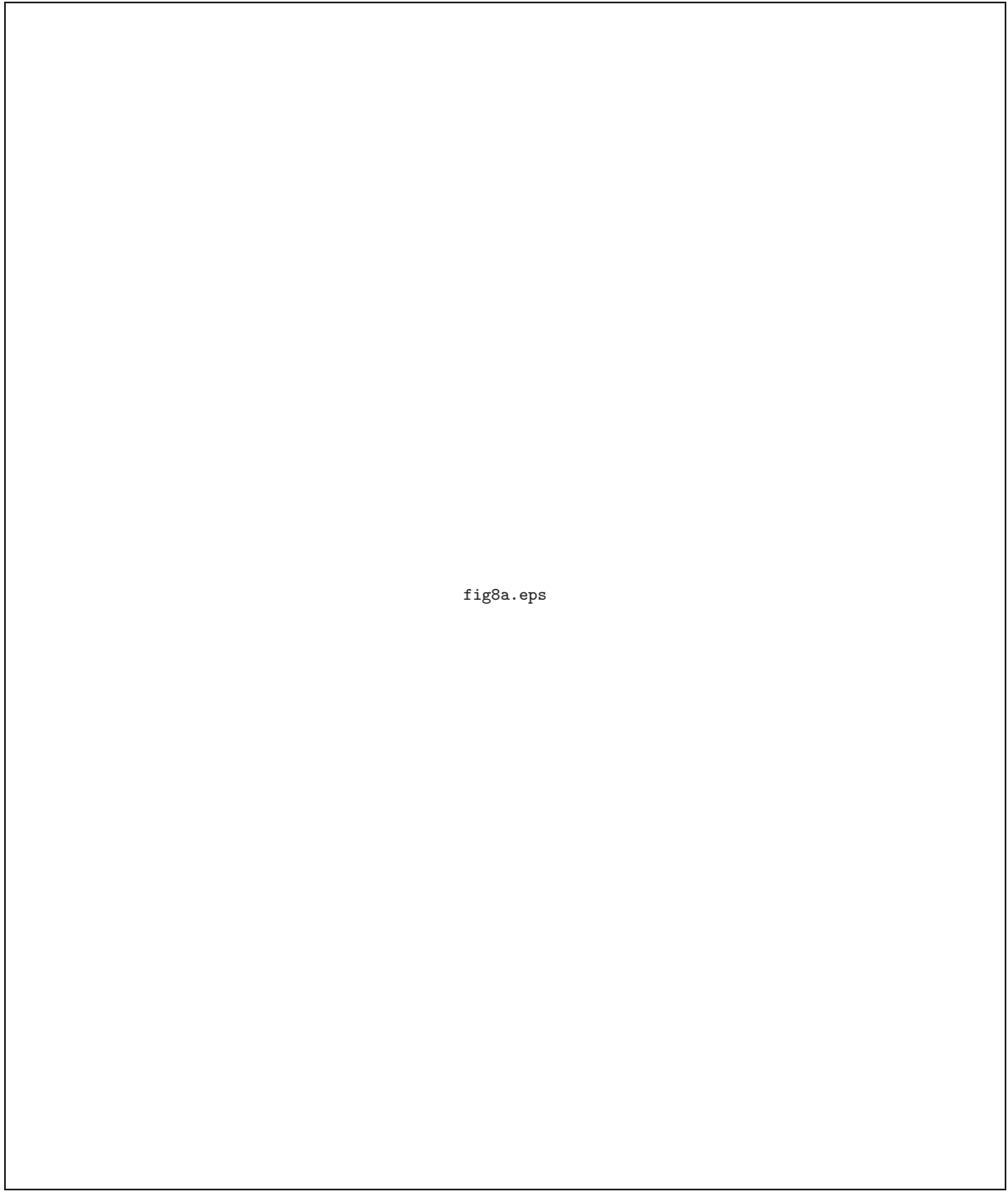

  \begin{center}
    \FigureFile(160mm,190mm){fig8a.eps}
  \end{center}
  \caption{a. Spectra of new SiO detections. The previous detections
  are not shown. 
  The 2MASS name (first half) and observed date (yymmdd.d) are shown on the upper left in each panel.
}\label{fig: spectra-a}
\end{figure*}
\setcounter{figure}{7}
\begin{figure*}
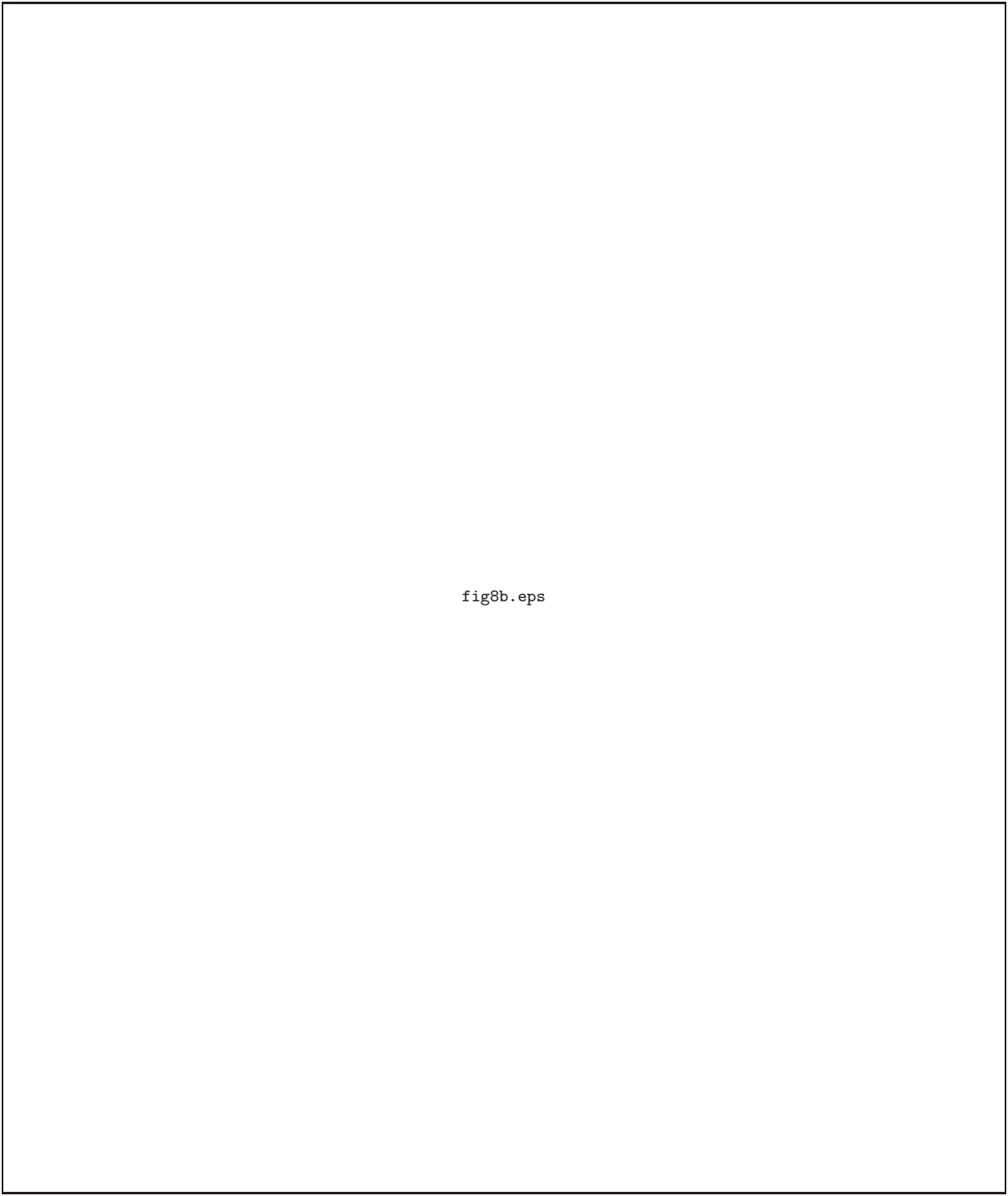

  \begin{center}
    \FigureFile(160mm,190mm){fig8b.eps}
  \end{center}
  \caption{b. continued
}\label{fig: spectra-b}
\end{figure*}
\setcounter{figure}{7}
\begin{figure*}
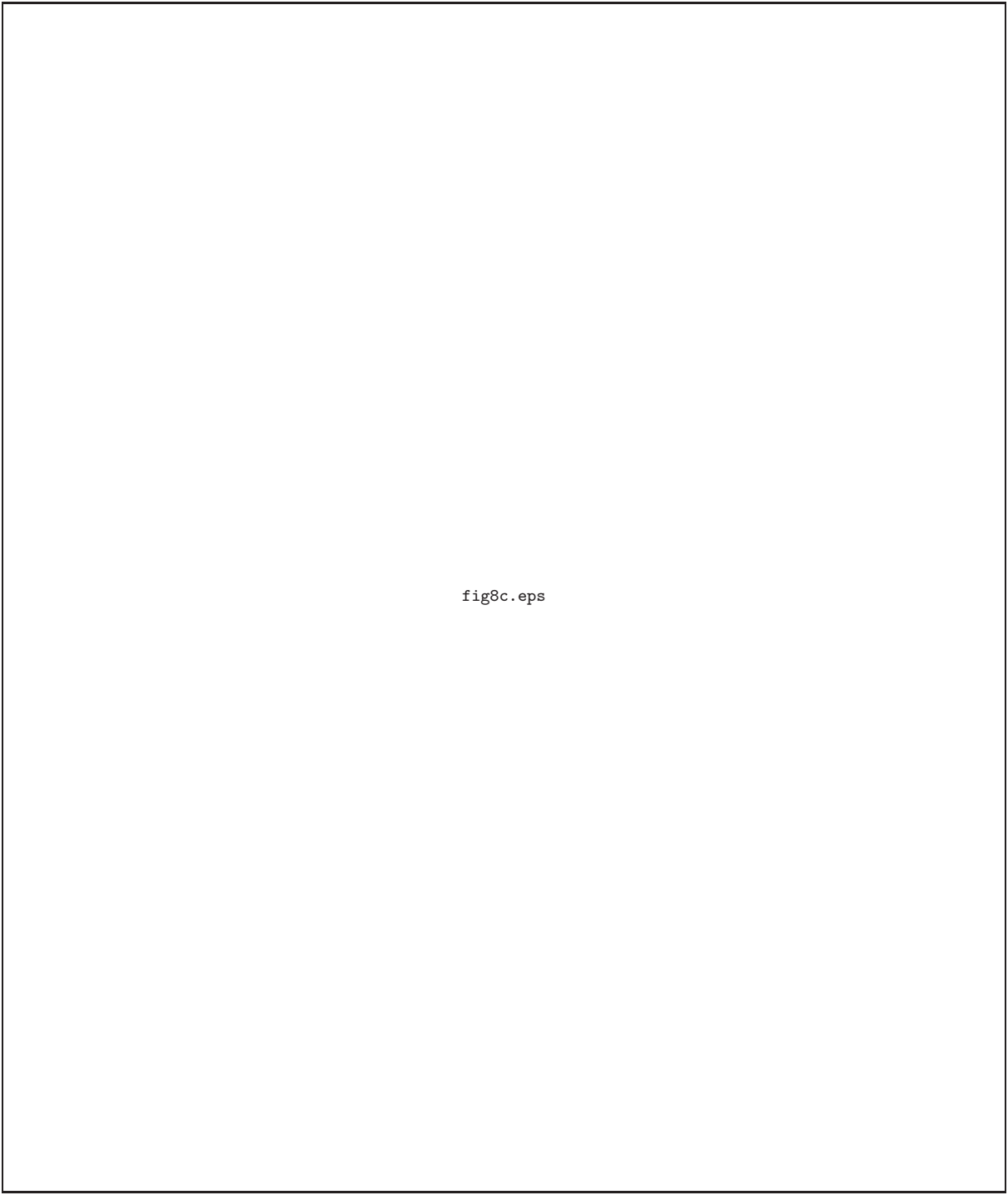

  \begin{center}
    \FigureFile(160mm,190mm){fig8c.eps}
  \end{center}
  \caption{c. continued
}\label{fig: spectra-c}
\end{figure*}
\setcounter{figure}{7}
\begin{figure*}
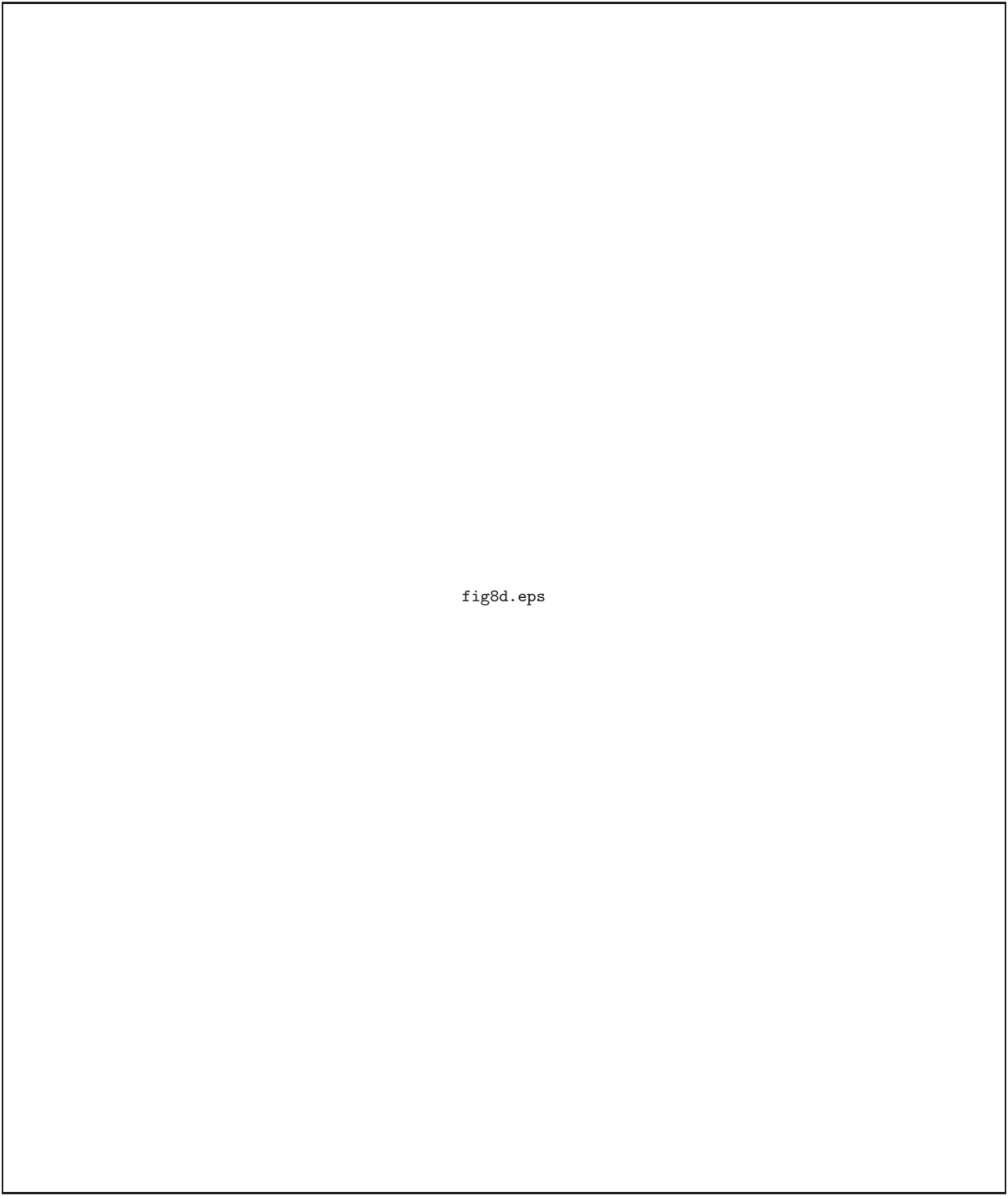

  \begin{center}
    \FigureFile(160mm,190mm){fig8d.eps}
  \end{center}
  \caption{d. continued
}\label{fig: spectra-d}
\end{figure*}
\setcounter{figure}{7}
\begin{figure*}
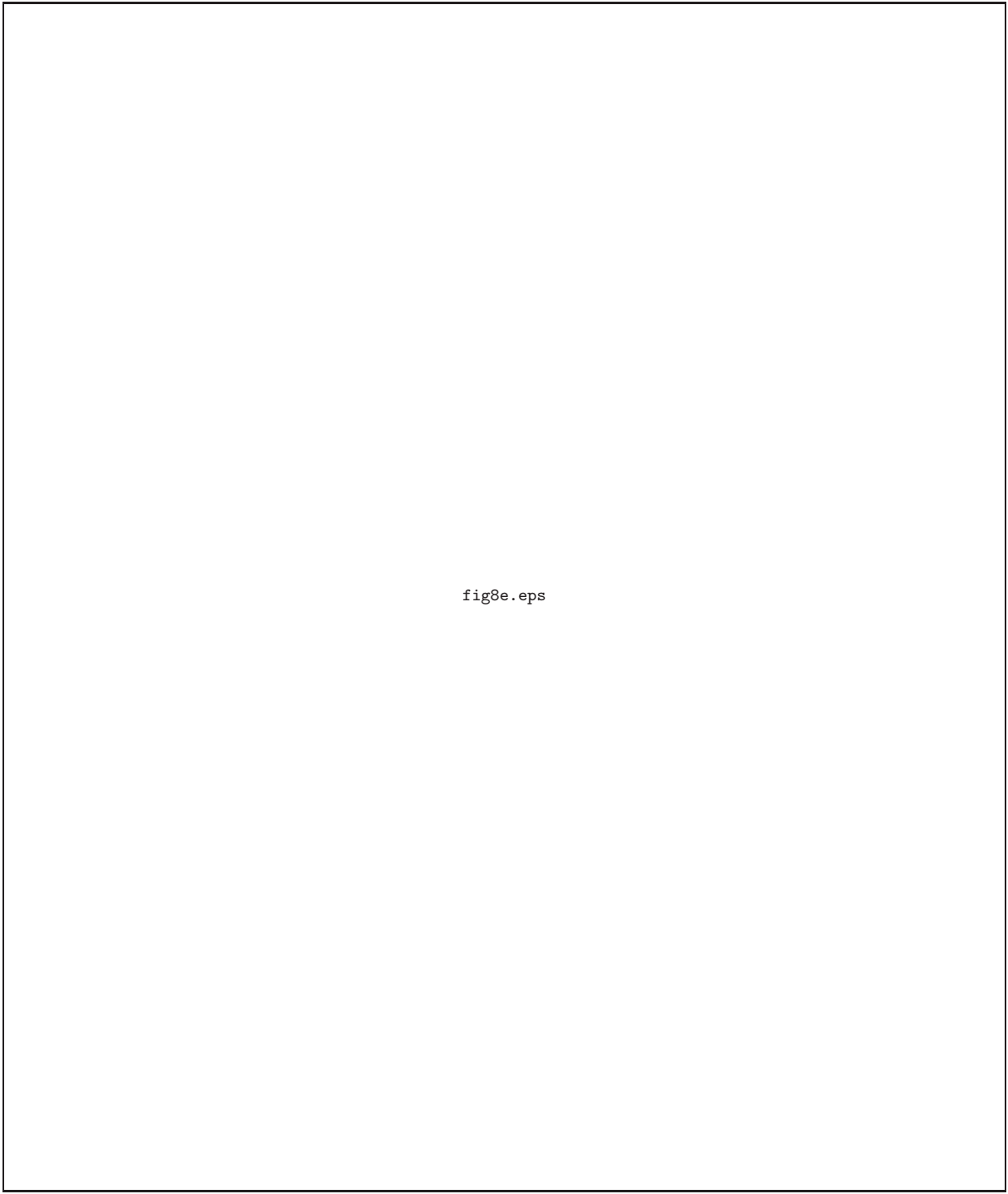

  \begin{center}
    \FigureFile(160mm,190mm){fig8e.eps}
  \end{center}
  \caption{e. continued
}\label{fig: spectra-e}
\end{figure*}
\setcounter{figure}{7}
\begin{figure*}
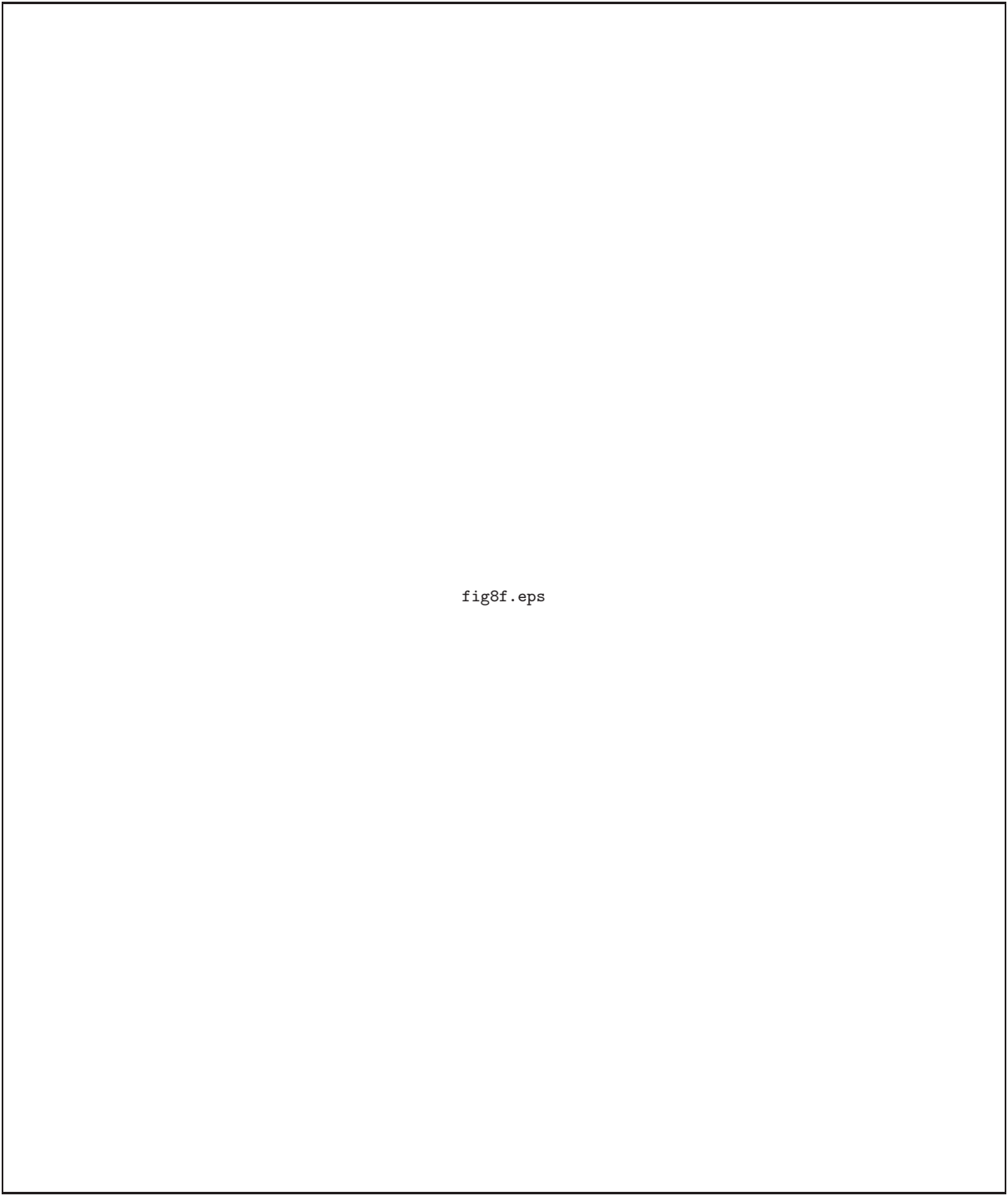

  \begin{center}
    \FigureFile(160mm,190mm){fig8f.eps}
  \end{center}
  \caption{f. continued
}\label{fig: spectra-f}
\end{figure*}
\setcounter{figure}{7}
\begin{figure*}
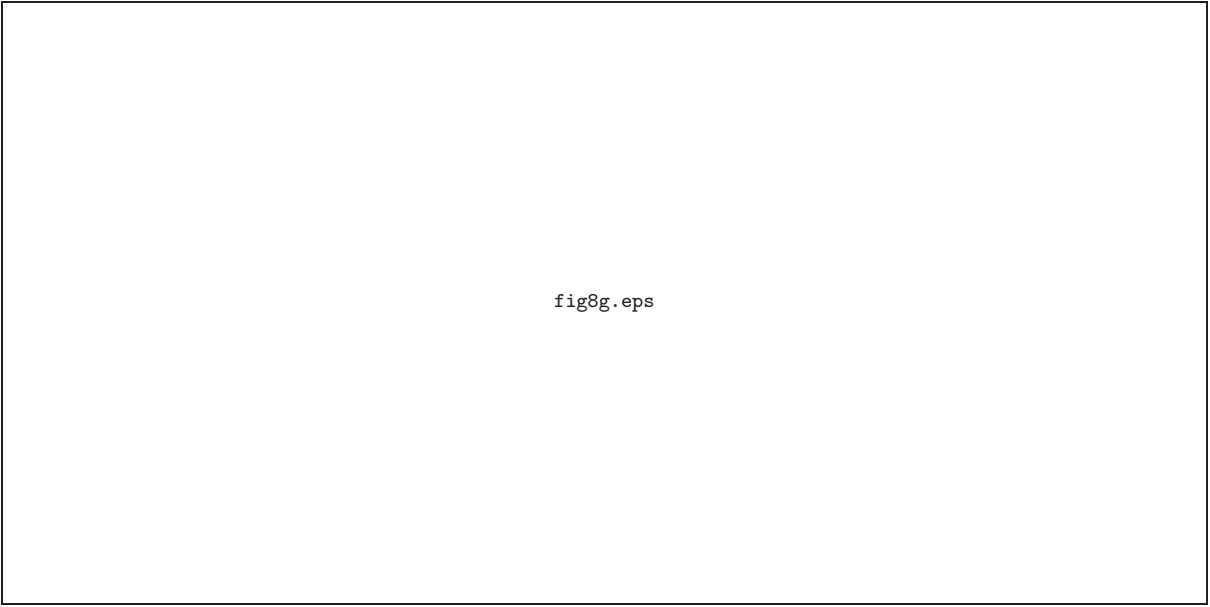

  \begin{center}
    \FigureFile(160mm,80mm){fig8g.eps}
  \end{center}
  \caption{g. continued
}\label{fig: spectra-g}
\end{figure*}

\begin{thebibliography}{}
\bibitem[Andronov et al.~(2006)]{and06}  Andronov, N., Pinsonneault, M. H., Terndrup, D. M.  2006, ApJ, 646, 1160
\bibitem[Bellazzini et al.~(2004) ]{bel04} Bellazzini, M., Ibata, R., Monaco, L., Martin, N., Irwin, M. J., \& Lewis, G. F.  2004, MNRAS, 354, 1263
\bibitem[Binney~(2001) ]{bin01} Binney, J. 2001, ASP Conf. Ser., 228, 269
\bibitem[Dame et al.(2001) ]{dam01}	Dame, T. M., Hartmann, D., \& Thaddeus, P. 2001,  \apj, 547, 792
\bibitem[David et al.~(1993)]{dav93} David, P., Le Squeren, A. M., Sivagnanam, P., Braz, M. A.1993, A\&AS, 98, 245
\bibitem[de Marchi et al.~(1999)]{dem99} de Marchi, G., Leibundgut, B., Paresce, F., \& Pulone, L.  1999, A\&A, 343, L9 
\bibitem[Deguchi et al.(1999)]{deg99} Deguchi, S., Fujii, T., Izumiura, H., Matsumoto, S., Nakada, Y., Wood, P. R., \& Yamamura, I. 1999, PASJ, 51, 355
\bibitem[Deguchi et al.(2000a)]{deg00a}  Deguchi, S., Fujii, T.,  Izumiura, H., Kameya, O.,  Nakada, Y., Nakashima, J., Ootsubo, T.,  \& Ukita, N.  2000a, \apjs, 128,  571
\bibitem[Deguchi et al.(2000b)]{deg00b}Deguchi, S., Fujii, T.,  Izumiura, H., Kameya, O., S. Nakada, Y., \& Nakashima, J. 2000b, \apjs,  130, 351
\bibitem[Deguchi et al.(2001)]{deg01} Deguchi, S.,  Nakashima, J., \&  Balasubramanyam,  R. 2004, \pasj,  53, 305
\bibitem[Deguchi et al.(2004a)]{deg04a} Deguchi, S., et al. 2004a, \pasj,  56, 261
\bibitem[Deguchi et al.(2004b)]{deg04b} Deguchi, S., et al. 2004b, \pasj,  56, 765
\bibitem[Deguchi et al.(2007)]{deg07} Deguchi, S.,  Nakashima, J., Kwok, S., \&  Konig,  N. 2007, \apj,  (submitted)
\bibitem[Djorgovski \&  Sosin~(1989) ]{djo89} Djorgovski, S. \&  Sosin, C. 1989, ApJ, 341, L13  
\bibitem[Eder et al.(1988)]{ede88} Eder, J., Lewis, B. M., \&  Terzian, Y. 1988, \apjs, 66, 183
\bibitem[Egan et al. (2003)]{ega03} Egan M.P., Price S.D., Kraemer K.E., Mizuno D.R., Carey S.J., Wright C.O., Engelke C.W., Cohen M., Gugliotti G. M. 2003, Air Force Research Laboratory Technical Report AFRL-VS-TR-2003-1589 available at {\it http://vizier.nao.ac.jp/viz-bin/VizieR?-source=V/114}
\bibitem[Fasano \& Franceschini~(1987) ]{fas87} Fasano, G. \& Franceschini, A. 1987, \mnras, 225, 155
\bibitem[Feast \& Whitelock~(2000)]{fea00} Feast, M. W. \& Whitelock, P. A. 2000, \mnras, 317, 460 
\bibitem[Fujii et al.(2006)]{fuj06} Fujii, T., Deguchi, S., Ita, Y., Izumiura, H., Kameya, O., Miyazaki, A., Nakada, Y. 2006, PASJ, 58, 529
\bibitem[Glass et al.(1995)]{gla95} Glass, I. S., Whitelock, P. A., Cathcpole, R. M. \&  Feast, M. W.  1995, MNRAS 273, 383
\bibitem[Harris(1996) ]{har96} Harris, W. E. 1996, \aj, 112, 1487
\bibitem[Helmi, White~(2001)]{hel01} Helmi, A. \& White, S. D. M. 2001, \mnras, 323, 529
\bibitem[Ibata et al.~(1994)]{iba94} Ibata, R. A., Gilmore, G., \& Irwin, M. J. 1994, Nature 370, 194   
\bibitem[Ibata et al.~(2001a)]{iba01a} Ibata, R., Irwin, M., Lewis, G., \& Stolte, A.  2001a, ApJ, 547, L133  
\bibitem[Ibata et al.~(2001b)]{iba01b} Ibata, R., Lewis, G., Irwin, M., Totten, E., \& Quinn, T. 2001b, ApJ, 551, 294 
\bibitem[Iben~(1981)]{ibe81} Iben, I., Jr. 1981, ApJ, 246, 278
\bibitem[Ita et al.(2001)]{ita01} Ita, Y., Deguchi, S., Fujii, T., Kameya, O., Miyoshi, M., Nakada, Y., Nakashima, J., \& Parthasarathy, M. 2001, \aap, 376, 112
\bibitem[Izumiura et al.(1995)]{izu95}Izumiura, H., Deguchi, S., Hashimoto, O., Nakada, Y., Onaka, T., Ono, T., Ukita, N.,  \& Yamamura, I. 1995, ApJ,   453,  837 
\bibitem[Jiang et al.(1996)]{jia96}Jiang, B. W., Deguchi, S.,  Yamamura,  I.,  Nakada, Y., Cho, S. H., \& Yamagata, T. 1996,  \apjs, 106, 463 
\bibitem[Jura et al.~(1993)]{jur93} Jura, M., Yamamoto, A., \& Kleinmann, S. G. 1993, ApJ, 413, 298
\bibitem[Kawamura et al.~(1999)]{kaw99} Kawamura, A., Onishi, T., Mizuno, A., Ogawa, H., \& Fukui, Y. 1999, PASJ, 51, 851
\bibitem[Kobulnicky et al.~(2005)]{kob05} Kobulnicky, H. A., Monson, A. J., Buckalew, B. A. et al. 2005, AJ, 129, 239 
\bibitem[Law et al.~(2005)]{law05}  Law, D. R., Johnston, K. V., \& Majewski, S. R.  2005, ApJ, 619, 807
\bibitem[Le Bertre et al.~(2001)]{leb01}	Le Bertre, T., Matsuura, M., Winters, J. M., Murakami, H., Yamamura, I., Freund, M., \& Tanaka, M. 2001, A\&A, 376, 997
\bibitem[Le Squeren et al.~(1992)]{les92} Le Squeren, A. M., Sivagnanam, P., Dennefeld, M., David, P. 1992, A\&A, 254, 133
\bibitem[Lewis~(1997)]{lew97} Lewis, B. M. 1997, AJ, 114, 1602
\bibitem[Lockwood et al.~(1985)]{loc85} Lockwood, G. W. 1985, ApJS, 58, 167 
\bibitem[Majewski et al.~(2003)]{maj03} Majewski, S. R., Skrutskie, M. F., Weinberg, M. D., \& Ostheimer, J. C. 2003, ApJ, 599, 1082
\bibitem[Majewski et al.~(2004)]{maj04} Majewski , S. R. et al. 2004, \apj, 128, 245
\bibitem[Martin et al.~(2004a) ]{mar04b} Martin, N. F., Ibata, R. A., Conn, B. C., Lewis, G. F., Bellazzini, M., Irwin, M. J., \& McConnachie, A. W.  2004, MNRAS, 355, L33 
\bibitem[Martin et al.~(2004b) ]{mar04a} Martin, N. F., Ibata, R. A., Bellazzini, M., Irwin, M. J., Lewis, G. F., \& Dehnen, W.  2004, MNRAS, 348, 12  
\bibitem[Martin et al.~(2005)]{mar05} Martin, N. F., Ibata, R. A., Conn, B. C., Lewis, G. F., Bellazzini, M., \& Irwin, M. J. 2005, \mnras, 362, 906
\bibitem[Matsunaga et al.~(2005)]{mat05} Matsunaga, N., Deguchi, S., Ita, Y., Tanabe, T., \& Nakada, Y. 2005, PASJ, 57, L1
\bibitem[Mauron et al.~(2005)]{mau05} Mauron, N., Kendall, T. R., \& Gigoyan, K. 2005, A\&A, 438, 867 
\bibitem[McWilliam \& Smecker-Hane~(2005)]{mcw05} McWilliam, A. \& Smecker-Hane, T. A. 2005, ASPC, 336, 221 Edited by Thomas G. Barnes III and Frank N. Bash, (Astronomical Society of the Pacific; San Francisco)
\bibitem[Messineo et al.(2002)]{mes02} Messineo, M., Habing, H. J., Sjouwerman, L. O., Omont, A., \& Menten, K. M. 2002, A\&A, 393, 115
\bibitem[Momany et al.(2006)]{mom06} Momany, Y., Zaggia, S., Gilmore, G., Piotto, G., Carraro, G., Bedin, L. R., \& de Angeli, F. 2006, A\&A, 451, 515
\bibitem[Navarro et al.~(2004)]{nav04} Navarro, J. F., Helmi, A., \& Freeman, K. C. 2004, \apjl, 601, L43
\bibitem[Nakashima,  Deguchi(2003)]{nak03} Nakashima, J., \& Deguchi, S. 2003, PASJ, 55, 203
\bibitem[Nishiyama et al.~(2006)]{nis06}  Nishiyama, S., Nagata, T., Kusakabe, N., et al. 2006, ApJ, 638, 839
\bibitem[Nyman et al.(1992) ]{nym92} Nyman, L.-\AA. Booth, R. S. Carlstr\"om, U. Habing, H. J. Heske, A. Sahai, R. Stark, R. van der Veen, W. E. C. J., \& Winnberg, A.1992,  \aaps, 93, 121
\bibitem[Nyman et al.(1998) ]{nym98} Nyman, L.-\AA., Hall, P.J., \& Olofsson, H. 1998, \aaps, 127, 185
\bibitem[Olnon et al.(1986)]{oln86} Olnon, F. M., Raimond, E. R., \&  IRAS Science team, 1986, \aaps, 65, 607
\bibitem[Paltrinieri et al.~(2001)]{pal01} Paltrinieri, B., Ferraro, F., Paresce, F., \& de Marchi, G. 2001, AJ, 121, 3114  
\bibitem[Press et al.~(1992) ]{pre92} Press, W. H., Flannery, B. P., Teukolsky, S. A., \& Vetterling, W. T. 1992, "Nemerical Recipe in C" (Cambridge University Press; Cambridge)
\bibitem[Sato (2002)]{sat02} Sato, S. 2002, The Proceedings of the IAU 8th Asian-Pacific Regional Meeting, Vol. II, Ed. by S. Ikeuchi, J. Hearnshaw, \& T. Hanawa, the Astronomical Society of Japan, p. 27
\bibitem[Sevenster et al.(1997)]{sev97} Sevenster, M. N., Chapman, J. M., Habing, H. J., Killeen, N. E. B., \& Lindqvist, M. 1997, \aaps, 122, 79
\bibitem[Sevenster et al.(2001)]{sev01} Sevenster, M. N., van Langevelde, H. J., Moody, R. A., Chapman, J. M., Habing, H. J., \& Killeen, N. E. B.  2001,  \aap, 366, 481
\bibitem[te Lintel-Hekkert et al.(1991)]{tel91} te Lintel-Hekkert, P., Caswell, J. L., Habing, H. J., Haynes, R. F., \& Norris, R. P. 1991, \aaps, 90, 327
\bibitem[van Loon et al.~(1996)]{van96} van Loon, J. T., Zijlstra, A. A.,  Bujarrabal, V., \& Nyman, L.-\AA. 1996, A\&A, 306, L29
\bibitem[Vig et al.~(2005) ]{vig05} Vig, S., Ghosh, S. K., \& Ojha, D. K.  2005, A\&A, 436, 867 
\bibitem[Wood et al.(1992)]{woo92} Wood, P. R.,Whiteark, S. M., Hughes. Sm. M. G, Bessel, M. S., Gardne. F. F., Hyland, A. R. 1992, ApJ, 397, 552
\bibitem[Yanny et al.~(2003)]{yan03} Yanny, B., Newberg, H.J., Grebel, E. K., Kent, S., et al. 2003, ApJ, 588, 824  
\end{thebibliography}
\end{document}